\documentclass[11pt, a4paper]{article}	

\usepackage[english]{babel}
\usepackage[T1]{fontenc}
\usepackage[utf8]{inputenc}
\usepackage{mathtools} 		
\usepackage{amssymb}  		
\usepackage{upgreek}
\usepackage{graphicx} 
\usepackage{comment}	
\usepackage[dvipsnames, x11names]{xcolor} 
\definecolor{Rosso}{cmyk}{0,1,0.8,0.2}	
\definecolor{Blu}{cmyk}{1,0.6,0,0.2}	

\usepackage{authblk}

\usepackage{cite}

\usepackage{hyperref}
  \hypersetup{linktoc=all, breaklinks, hidelinks, colorlinks=true, linkcolor=Blu, citecolor=Blu, urlcolor=Blu,}
\usepackage{bookmark} 

\newcommand*{\iu}{\mathrm{i}\mkern1mu}  
\newcommand*{\e}{\mathrm{e}} 
\newcommand*{\mat}[1]{\begin{pmatrix} #1 \end{pmatrix}}
\newcommand*{\vb}[1]{\boldsymbol{#1}}
\newcommand*{\dd}[1]{\mathrm{d}#1}
\newcommand*{\dom}{\mathfrak{D}} 
\newcommand*{\Leb}{L} 
\newcommand*{\Sob}{H} 
\newcommand*{\N}{\mathbb{N}}
\newcommand*{\Z}{\mathbb{Z}}	
\newcommand*{\R}{\mathbb{R}}

\newcommand*{\UU}{\mathrm{U}}	
\newcommand*{\SU}{\mathrm{SU}}
\DeclareMathOperator{\sign}{sgn}
\DeclareMathOperator{\sinc}{sinc}
\DeclareMathOperator{\tr}{tr}

\DeclareMathOperator{\id}{id}
\DeclareMathOperator{\Inn}{Inn}
\DeclareMathOperator{\Out}{Out}
\DeclareMathOperator{\Aut}{Aut}

\DeclareMathOperator{\spec}{spec}

\newcommand*{\g}{\vb{g}}
\newcommand*{\m}{\vb{m}}
\newcommand*{\ddelta}{\updelta}
\newcommand*{\dk}{\dd k}
\renewcommand*{\Gamma}{\varGamma}
\renewcommand*{\Delta}{\varDelta}
\renewcommand*{\Psi}{\varPsi}
\renewcommand*{\Phi}{\varPhi}

\newcommand*{\UB}{\mathcal{U}_{\text{B}}}

\title{Isospectrality and non-locality of generalized Dirac combs}

\date{}

\author[$\hspace{0cm}$]{Giuliano Angelone$^{1,}$\footnote{Email address: \href{mailto:giuliano.angelone@uniroma1.it}{giuliano.angelone@uniroma1.it} }}
\author[$\hspace{0cm}$]{Manuel Asorey$^{2}$}
\author[$\hspace{0cm}$]{Fernando Ezquerro$^{2}$}
\author[$\hspace{0cm}$]{Paolo Facchi$^{3,4}$}

\affil[1]{Dipartimento di Matematica, Sapienza Università di Roma, 00185 Rome, Italy}
\affil[2]{Centro de Astropart\'{i}culas y F\'{i}sica de Altas Energ\'{i}as, 
Departamento de F\'{i}sica Te\'{o}rica, Universidad de Zaragoza, 50009 Zaragoza, Spain}
\affil[3]{Dipartimento di Fisica, Universit\`{a} di Bari, 70126 Bari, Italy}
\affil[4]{INFN, Sezione di Bari, 70126 Bari, Italy}

\begin{document}
\maketitle

\begin{abstract}
We consider a generalization of  Dirac's comb model, describing a non-relativistic particle moving in a periodic array of generalized point interactions. The latter represent the most general point interactions rendering the kinetic-energy operator self-adjoint, and form a four-parameters family that includes the $\ddelta$-potential and the $\ddelta'$-potential as particular cases. We study the parameter dependence of the spectral properties of this system, finding a rich isospectrality structure. We systematically classify a large class of isospectral relations, determining which Hamiltonians are spectrally unique, and which are instead related by a unitary or anti-unitary transformation.
\end{abstract}

\section{Introduction}
The analysis of quantum systems with singular interactions has over the years led to a wide class of solvable models that, despite their apparent simplicity, capture  the essential features of more complex systems in a mathematically tractable form~\cite{Yang, AGHH88, AlbKur00}. Singular interactions have been extensively used, among other applications, to describe atomic systems~\cite{DemOst88}, condensed matter systems with localized structure defects~\cite{Gal91}, to study the properties of nanotubes~\cite{AdTi07}, and to characterize the Casimir effect in the presence of metamaterials boundaries~\cite{AsoMun13, MuMaMo13, MCandJMG}.
Of particular relevance in this context are point interactions, a powerful tool that allows to model quantum systems with very short-range interactions located around a given point, providing deep insights into phenomena such as scattering, localization, tunneling, and band formation. These idealized zero-range interactions are usually described either in terms of $\ddelta$-like potentials or effective boundary conditions, but both descriptions can be formalized within the rigorous framework of self-adjoint extensions of symmetric operators.

Even in one dimension, this theory exhibits a rich phenomenology: the kinetic-energy operator of a spinless particle (that is, the Laplacian), restricted to functions with a discontinuity or kink at a point, admits a four-parameter family of self-adjoint extensions. The latter characterizes a large variety of physical systems involving $\ddelta$- and $\ddelta'$-potentials, as well as more general interactions which can violate parity, time-reversal, or induce non-local effects. The physical description of this four-parameter model involves, beside localized potentials, also mass-jumps and localized magnetic fluxes~\cite{KuPa15,KuPa19}. From an experimental perspective the corresponding Hamiltonians can be realized by using Josephson junctions and other quasi one-dimensional heterogeneous structures where the magnetic flux can be controlled in the transition layer~\cite{BaPa82}.

A paradigmatic example of a one-dimensional periodic system containing point interactions is the well-known Kronig--Penney model~\cite{kronig}, sometimes referred to as a \emph{Dirac comb}, which consists in the periodic arrangement of $\ddelta$-potentials. 
This system stands as a minimal, yet non-trivial, prototype for the study of band structures, Bloch theory and spectral properties of crystals. For this reason, it has inspired a substantial literature that analyzes how different types of point interactions affect spectral properties such as band gaps, embedded eigenvalues, and localization~\cite{GesHol87, Hug98, ExGr99, KuLa02, CheShi04, Bordag19}. Extensions of the Kronig--Penney model to include $\ddelta'$ interactions, for instance, have revealed novel phenomena like non-symmetric dispersion relations and unusual spectral effects. 

In this paper, we study a generalization of the Dirac comb by considering the most general periodic point interaction which gives rise to a self-adjoint extension of the kinetic-energy operator. We parametrize the point interactions via unitary matrices in $\UU(2)$, encoding thus the full four-parameters family of self-adjoint extensions. This parametrization unifies the treatment of various types of point interactions, but also facilitates the study of the spectral properties of the system. As it turns out, the spectrum depends only on a subset of these parameters, revealing an unexpected \emph{isospectrality}: distinct physical systems, associated with distinct point interactions, can share the same spectrum. Our main contribution is to provide a systematic classification of the isospectral structure by considering a large set of isospectral transformations, including both unitary and antiunitary transformations, some of which act non-locally in the bulk. 
Notice that, more generally, the analysis of the isospectrality structure can shed light on other inverse spectral problems, helping to understand the extent to which spectral data encode the geometrical and physical properties of the system~\cite{Kac, Zel04}. Moreover, similar generalizations of the Kronig--Penney model have been recently considered in order to construct simple systems displaying non-trivial topological phases~\cite{ReBeRo19, SmiPri20}.

The paper is organized as follows. In Sec.~\ref{sec:genpointint}, we review the theory of generalized point interactions in one dimension, both from the perspective of unitary boundary conditions and of singular potential realizations. In Sec.~\ref{sec:genComb} we introduce the generalized Dirac comb and derive its spectral function via a Bloch decomposition. In Sec.~\ref{sec:isospectrality} we extensively discuss the isospectral structure of the system, identifying  many spectral symmetries and discussing some of their physical implications. We conclude in Sec.~\ref{sec:discussion} with a discussion of the broader significance of our results and possible extensions.

\section{Generalized point interactions in one dimension}\label{sec:genpointint}
Consider a nonrelativistic spinless quantum particle subjected to a \emph{generalized point interaction} in dimension one, that is to the most general perturbation, supported at a single point (say, the origin $x=0$), preserving the self-adjointness of the kinetic-energy operator
\begin{align}\label{eq:H0}
H_0=-\frac{\hbar^2}{2m}\frac{ \dd^2}{\dd x^2}\,,
\end{align}
namely the one-dimensional Laplacian. As is well known,  when initially defined on $C_0^\infty(\R\! \setminus \!\{0\})$, the space of smooth functions with compact support on the punctured line, the above operator is symmetric but not self-adjoint, and the full set of self-adjoint extensions form a four-parameter family. Remarkably, the set of self-adjoint extensions of $H_0$ coincides with the set of its zero-range perturbations: depending on the context, the self-adjoint extensions of $H_0$ have been  in fact described either in terms of singular (i.e.\ $\ddelta$-like) point interactions~\cite{AGHH88, Kur96, ExGr99, AlbKur00, BrzJef01,  HaLaPa05, Lan15, DeVSan16} or in terms of suitable boundary conditions to be satisfied by the wave function~\cite{AkhGla81, BFV01, AIM05, AIM15, ALPP15, FGL18}. 
In one dimension, the two descriptions are substantially interchangeable, and offer complementary perspectives on the physical system under investigation. For this reason, after briefly reviewing in sections~\ref{sec:ubc} and~\ref{sec:pointint} both formalisms, explicitly clarifying how they are related, we devolve section~\ref{sec:examples} to present many particular cases that are  physically interesting. Before proceeding let us remark that all the self-adjoint extension correspond to the same physics in the ``bulk'', that is far from the origin, but they describe different interactions between the particle and the ``boundary'', giving thus rise to different global dynamics and represent different physical systems: the origin can be thought as a junction, or a defect, whose physical properties are encoded in the choice of the self-adjoint extension.

\subsection{Unitary boundary conditions}\label{sec:ubc}
We start by the ``unitary'' boundary conditions approach, which is intimately related to the problem of finding the self-adjoint extensions of $H_0$. All self-adjoint realizations of the operator in~\eqref{eq:H0} are known to be in one-to-one correspondence with the set of $2\times 2$ unitary matrices $U\in\UU(2)$~\cite{AIM05,AIM15}. Each of these realizations, which we henceforth denote by $H_{0,U}$, is defined on the domain
\begin{equation}\label{eq:PsiPsiprime}  
\dom(H_{0,U})=\bigl\{\psi\in\Sob^2(\R\!\setminus\!\{0\}): (I-U)\Psi_0'=\iu (I+U)\Psi_0\bigr\}
\end{equation}
where $\Sob^2(\Omega)$ is the space of functions with square-integrable first and second weak derivative (that is, the second Sobolev space) with support in $\Omega$, $I$ is the $2\times 2$ identity matrix whereas\footnote{It is worth mentioning that any wave function $\psi(x)$ in $\Sob^2(\R\!\setminus\!\{0\})$ is continuous and differentiable at any $x\neq0$ but can be discontinuous at $x=0$, and one has thus to distinguish the left and right limits of both $\psi(x)$ and $\psi'(x)$}.
\begin{align}
\Psi_0=\mat{  \psi(0^-)\\  \psi(0^+)} \,, && \Psi_0'=\ell\mat{ \psi'(0^-) \\ -\psi'(0^+)} 
\end{align}
are two vectors containing, respectively, the boundary values of $\psi(x)$ and of its normal derivative at $x=0$, with $\ell$ being an arbitrary reference length needed for dimensional reasons. For each $U\in \UU(2)$ the equation 
\begin{align}\label{eq:UBC}
(I-U)\Psi_0'=\iu (I+U)\Psi_0
\end{align}
describes thus a particular  boundary condition,  to be satisfied by the wave function at the origin, that ensures the self-adjointness of $H_{0,U}$. In the following we parametrize a generic $\UU(2)$ matrix as in~\cite{isoboundary},
\begin{align}\label{eq:U}
	U=U(\eta,\vb{m})=\e^{\iu\eta}\mat{m_0+\iu m_3 & m_2+\iu m_1 \\ -m_2+\iu m_1 & m_0-\iu m_3}\,,
\end{align}
where $\eta\in[0,\pi)$ and $\m=(m_0,m_1,m_2,m_3)\in\R^4$ with $\|\m\|^2=1$.

The parametrization of the boundary conditions in terms of a unitary matrix via Eq.~\eqref{eq:UBC} is quite natural from the point of view of the self-adjoint extensions, and as we will show is also very useful to determine the spectral properties of the Hamiltonian, but it is not unique. 
Another choice that is well suited for our scopes~\cite{ExGr99, Lan15} is to express the boundary conditions in terms of the  \emph{jump} $[\psi(0)]$  and the \emph{average} $\{\psi(0)\}$ of $\psi(x)$ (and of its derivative) at $x=0$, defined as
\begin{align}
\{\psi(0)\}=\frac{\psi(0^+)+\psi(0^-)}{2}\,,&&[\psi(0)]=\psi(0^+)-\psi(0^-)\,.
\end{align}
The corresponding jump-average boundary conditions, that we introduce in the next section, naturally arise when the self-adjoint extensions of $H_0$ are realized by means of suitable point interactions supported at the origin.

\subsection{Point interactions}\label{sec:pointint}
We now consider the operator that is obtained by adding to $H_0$ the Kurasov (pseudo-)potential $V_{\g}(x)$, namely
\begin{align}\label{eq:H0g}
\tilde{H}_{0,\g}=-\frac{\hbar^2}{2m}\frac{ \dd^2}{\dd x^2}+V_{\g}(x)\,,&&V_{\g}(x)
=\sum_{i=1}^{4} g_i\, v_i(x)\,,
\end{align}
where $\g=(g_1,g_2,g_3,g_4)$ is a vector containing the dimensionless coupling constants of the following  point interactions supported at the origin:
\begin{subequations}
\begin{gather} 
	v_1(x)=\frac{\hbar^2}{m\ell} \ddelta(x) \,,\qquad
	v_2(x) 
	=\frac{\hbar^2}{m} \ddelta'(x)\,,\\[2pt]
	v_3(x) 
    =\iu \frac{\hbar^2}{m} \biggl(\frac{\dd}{\dd x}\ddelta(x)+\ddelta(x)\frac{\dd}{\dd x}\biggr)\,,\qquad 
	v_4(x) 
	=\frac{\hbar^2\ell}{m} \frac{\dd}{\dd x}\ddelta(x)\frac{\dd}{\dd x}\,.
	\label{eq:v3v4def}
\end{gather}
\end{subequations}
The Hamiltonian $\tilde{H}_{0,\vb{g}}$ has to be properly interpreted in order to represent a well-defined (self-adjoint) operator on the Hilbert space $\Leb^2(\R)$, as the Dirac delta distribution $\ddelta(x)$ and its derivative $\ddelta'(x)$ are not multiplication operators on $\Leb^2(\R)$ and, in the context of standard distribution theory~\cite{Hor03}, their  product with a non-smooth function is not defined. A rigorous description of  $\tilde{H}_{0,\g}$  requires the formalism of finite rank perturbations via a suitable scale of Hilbert spaces, and to generalize distribution theory to discontinuous functions~\cite{Kur96, AlbKur00, Lan15}. 

At the very end, however, the action of $\tilde{H}_{0,\g}$ is equivalent to that of the free Hamiltonian $H_0$ on $H^2(\R\!\setminus\!\{0\} )$ supplied with the following jump-average condition at the origin:
\begin{align}\label{eq:jumpaverageBC}
\mat{ \ell [\psi'(0)]\\[2pt] [\psi(0)]}=2M_{\vb{g}} \mat{ \{\psi(0)\}\\[2pt] \ell \{\psi'(0)\}}\,,&&   M_{\vb{g}} =\mat{g_1& -g_2+\iu g_3\\ g_2+\iu g_3 &g_4}\,.
\end{align}
In particular, if $\g\in\R^4$, i.e.\ if the $g_i$'s are real numbers, the above boundary conditions parametrize \emph{most} self-adjoint extensions of $H_0$. 
The full set is recovered by considering the vector $\g$ as an element of the real projective space $\R\mathbb{P}^4$, that is by allowing the  $g_i$'s to  take infinite values, see e.g.\ Theorem~3.2.7 of~\cite{AlbKur00}. This fact can be substantiated by expressing the elements of the matrix $M_{\vb{g}}$ in terms of the elements of the unitary matrix $U(\eta, \vb{m})$, that is by determining the relation $\vb{g}=\vb{g}(U)$:
\begin{align}\label{eq:gU}
\mat{g_1\\ g_2\\g_3\\g_4}=
\frac{1}{m_1+\sin(\eta)}\mat{m_0+\cos(\eta)\\-m_3\\m_2\\m_0-\cos(\eta)}\,,
\end{align}
see appendix~\ref{app:Kurasov}, and~\cite{Lan15}, for further details.  As anticipated, one should formally consider infinite values of the $g_i$'s when $m_1+\sin(\eta)=0$. In any case, we can conclude that for any $U\in\UU(2)$  there exists a quadruple of real (and possibly infinite) parameters $\vb{g}=\vb{g}(U)$ such that the action of $H_{0,U}$ coincides with the action of $\tilde{H}_{0,\g(U)}$. We will henceforth refer to this relation  as the \emph{Kurasov mapping}. 

Despite the existence of this mapping, we stress that the differential expression in~\eqref{eq:H0g} has a physical relevance on its own, as the singular potential $V_{\g}(x)$ can be approximated by a suitable family of smooth (and thus well-defined) potentials~\cite{Kur96}. Moreover, the four singular interactions entering in the expression of $V_{\g}(x)$ have a clear physical interpretation, as explained in the following. The  potentials $v_1(x)$ and $v_2(x)$  are associated with two different zero-range interactions, described by the $\ddelta$- and $\ddelta'$-potentials:
\begin{align}
\tilde{H}_{0,(g_1,g_2,0,0)}=-\frac{\hbar^2}{2m}\frac{ \dd^2}{\dd x^2}+\frac{\hbar^2}{m}\biggl(\frac{1}{\ell}g_1\ddelta(x)+g_2\ddelta'(x)\biggr)\,.
\end{align}
The pseudo-potential $v_3(x)$ is associated with a (regularized) singular gauge field, as $\tilde{H}_{0,(0,0,g_3,0)}$ heuristically corresponds to the operator
\begin{align}
-\frac{\hbar^2}{2m}\biggl(\frac{ \dd}{\dd x}-2\iu g_3\ddelta(x)\biggr)^2-\frac{\hbar^2}{2m}\bigl(2g_3\ddelta(x)\bigr)^2\,.
\end{align}
 The pseudo-potential $v_4(x)$, finally, resembles a one-dimensional Laplace--Beltrami operator  with a singular metric:
\begin{equation} 
\tilde{H}_{0,(0,0,0,g_4)} =-\frac{\hbar^2}{2m}\frac{ \dd}{\dd x}\bigl(1-2\ell g_4\ddelta(x)\bigr)\frac{ \dd}{\dd x}\,.
\end{equation}

\subsection{Examples}\label{sec:examples}
In this section look at  examples of boundary conditions and singular point interactions, determining their mutual connection.

\subsubsection{Robin boundary conditions}
The unitary matrix
\begin{align}
    U=\mat{\e^{\iu (\eta-\alpha)}& 0 \\ 0 &\e^{\iu(\eta+\alpha)}}
\end{align}
corresponding to the parameters
\begin{align}
\eta\in[0,\pi)\,,&&\m=\bigl( \cos(\alpha), 0, 0, -\sin(\alpha)\bigr)\,,
\end{align}
for $\alpha\in[0,2\pi)$, leads to the asymmetric Robin conditions
\begin{align}\label{eq:URobin}
    \psi'(0^\pm)=\pm\frac{1}{\ell}\cot\Bigl(\frac{\alpha_\pm}{2}\Bigr)\psi(0^\pm)\,,
\end{align}
where  $\alpha_\pm=\eta\pm\alpha$. These are the most general boundary conditions ensuring that the probability current density associated with a wave function $\psi\in H^2(\R)$,
\begin{align}
    j_\psi(x)=-\frac{\hbar^2}{2m}\Bigl(\overline{\psi'(x)}\psi(x)-\overline{\psi(x)}\psi'(x)\Bigr)\,,
\end{align}
  vanishes at the origin, i.e.\ $j_\psi(0^+)=j_\psi(0^-)=0$. For this reason, they are denoted as confining, or separated, boundary conditions: a wave function initially  
supported in (say) the left half-line $(-\infty,0)$ will never cross the origin throughout the evolution generated by the corresponding Hamiltonian $H_{0,U}$. 
When $\e^{\iu\alpha_+}=\e^{\iu \alpha_-}$, so that $\alpha\in\{0,\pi\}$,  they reduce to the symmetric Robin conditions, parametrized by 
\begin{align}\label{eq:alpha0}
\eta\in[0,\pi)\,,&&\m=(1,0,0,0),
\end{align}
if $\alpha=0$ and 
\begin{align}\label{eq:alphapi}
\eta\in[0,\pi)\,,&& \m=(-1,0,0,0),  
\end{align}
if $\alpha=\pi$. For $\eta=0$ Eqs.~\eqref{eq:alpha0} and~\eqref{eq:alphapi} respectively give the well-known Dirichlet and Neumann boundary conditions, 
\begin{align}
    \psi(0^-)=\psi(0^+)=0\,,&&\psi'(0^-)=\psi'(0^+)=0\,.
\end{align}

In terms of the couplings $\vb{g}$ characterizing the jump-average boundary conditions the asymmetric Robin condition are obtained for
\begin{align} 
    \g=\biggl(
    \frac{\cos(\alpha)+\cos(\eta)}{\sin(\eta)},
    \frac{\sin(\alpha)}{\sin(\eta)},
    0,
    \frac{\cos(\alpha)-\cos(\eta)}{\sin(\eta)}
    \biggr)
\end{align}
whereas the symmetric Robin conditions are obtained for
\begin{align}
    \g=\Bigl(\cot\Bigl(\frac{\eta+\alpha}{2}\Bigr), 0, 0,  \tan\Bigl(\frac{\eta+\alpha}{2}\Bigr)\Bigr)
\end{align}
and correspond thus to a coupled   $(g_1,0,0,g_4)$ singular interaction.

\subsubsection{Pseudo-periodic boundary conditions}\label{sec:ppbc}
The unitary matrix
\begin{align}
    U=\mat{0& -\e^{-\iu\alpha} \\ -\e^{\iu\alpha}&0}
\end{align}
corresponding to the parameters
\begin{align} 
 \eta=\frac{\pi}{2}\,,&&  \m=\bigl(0, \cos(\alpha), \sin(\alpha),0 \bigr)\,,
\end{align}
for $\alpha\in[0,2\pi)$, leads to the pseudo-periodic conditions
\begin{align}\label{eq:Upp}
    \psi(0^+)=\e^{\iu\alpha}\psi(0^-)\,,&& \psi'(0^+)=\e^{\iu\alpha}\psi'(0^-)\,,
\end{align}
reducing to the periodic and anti-periodic conditions respectively for $\alpha=0$ and $\alpha=\pi$. With these boundary conditions the probability current density does not vanishes at the origin, and thus they are  denoted as non-confining, or connected. Notice that for each non-confining boundary condition the outgoing current  is exactly balanced by the incoming current, namely
$j_\psi(0^-) = j_\psi(0^+)$, in accordance with the unitarity of the evolution and the conservation of probability ensured by the self-adjointness of the Hamiltonian.

Pseudo-periodic conditions can be implemented by the couplings
\begin{align}
  \g=\Bigl(0,0, \tan\Bigl(\frac{\alpha}{2}\Bigr), 0\Bigr) 
\end{align}
and correspond thus to the most general $(0,0,g_3,0)$ singular interaction.

\subsubsection{Imaginary quasi-periodic boundary conditions}
The unitary matrix
\begin{align} 
    U=\mat{-\cos(\alpha)& \iu\sin(\alpha) \\ -\iu \sin(\alpha) & \cos(\alpha)},
\end{align}
corresponding to the parameters
\begin{align}
\eta=\frac{\pi}{2}\,,&&\m=\bigl(0,0,\sin(\alpha),\cos(\alpha)\bigr),
\end{align}
for $\alpha\in[0,2\pi)$, leads to the imaginary quasi-periodic conditions
\begin{align}\label{eq:imqpbc}
    \psi(0^+)=\iu\tan\Bigl(\frac{\alpha}{2}\Bigr)\psi(0^-)\,,&& \psi'(0^+)=\iu\cot\Bigl(\frac{\alpha}{2}\Bigr)\psi'(0^-)\,,
\end{align}
which reduce to the mixed Neumann-Dirichlet (or Zaremba) conditions
\begin{align}
    \psi'(0^-)=0\,,&& \psi(0^+)=0
\end{align}
for $\alpha=0$ and to the the pseudo-periodic conditions with phase $\e^{\iu \pi/2}$ for $\alpha=\pi/2$. As derived in~\cite{parities}, but see also~\cite{isoboundary, isodirac}, these quasi-periodic boundary conditions constitute an example of isospectral boundary conditions for this system, since the corresponding Hamiltonians are found to be unitarily equivalent.

Quasi-periodic conditions can be implemented by the couplings
\begin{align}
    \g=\bigl(0,- \cos(\alpha), \sin(\alpha),0\bigr)
\end{align}
corresponding thus to a coupled $(0,g_2,g_3,0)$ singular interaction.

\subsubsection{Singular interactions}
We conclude this section by determining which boundary conditions correspond to the (uncoupled) singular interactions. For this scope we invert the non-linear relations in~\eqref{eq:gU}, obtaining
\begin{subequations}
\begin{gather}
\eta=\arctan\biggl(\frac{\Delta}{g_1-g_4}\biggr) \bmod \pi \,,\\
\mat{m_0\\ m_1\\m_2\\m_3}=
\frac{\sign(\Delta)}{\sqrt{(g_1-g_4)^2+\Delta^2}}\mat{g_1+g_4\\2-\Delta\\2g_3\\-2g_2}
\end{gather}
\end{subequations}
if $\Delta=1+g_1g_4+g_2^2+g_3^2\neq 0$, and
\begin{subequations}
\begin{gather}
\eta=0\,,\\
\mat{m_0\\ m_1\\m_2\\m_3}=
\frac{1}{g_1-g_4}\mat{g_1+g_4\\\-2\\2g_3\\-2g_2}
\end{gather}
\end{subequations}
if $\Delta=0$. Notice that the quantity $(g_1-g_4)^2+\Delta^2$ never vanishes, and is thus strictly positive, and similarly $g_1-g_4$ cannot vanish when $\Delta=0$. 

By using the above formulae we find the following relations. The singular interaction $\g=(g_1,0,0,0)$ corresponds to the jump-average conditions
\begin{align}
\psi(0^+)=\psi(0^-)\,&& \ell\bigl(\psi'(0^+)-\psi'(0^-)\bigr)= 2 g_1 \psi(0^+)
\end{align}
and is realized by the parameters
\begin{align}\label{eq:g1}
\eta=\frac{\pi}{2}-\arctan(g_1)\,,&&\m=\biggl( \frac{g_1}{\sqrt{1+g_1^2}}, \frac{1}{\sqrt{1+g_1^2}},0,0\biggr)
\end{align}
determining the unitary matrix $U=U(\eta,\vb{m})$. The interaction $\g=(0,g_2,0,0)$ corresponds to the \emph{real} quasi-periodic conditions 
\begin{align}
\psi(0^+)=\frac{1+g_2}{1-g_2}\psi(0^-)\,,&&
\psi'(0^+)=\frac{1-g_2}{1+g_2}\psi'(0^-)\,,
\end{align}
not to be confused with the imaginary quasi-periodic conditions in~\eqref{eq:imqpbc}, and is realized by the parameters
\begin{align}\label{eq:g2}
\eta=\frac{\pi}{2} \,, &&\m=\biggl(0, \frac{1-g_2^2}{1+g_2^2}, 0,  \frac{-2g_2}{1+g_2^2},\biggr)\,.
\end{align}
The interaction $\g=(0,0,g_3,0)$ is equivalent to the pseudo-periodic conditions, as already found in section \ref{sec:ppbc}. The interaction $\g=(0,0,0,g_4)$, finally, corresponds to the jump-average conditions
 \begin{align}
\psi(0^+)-\psi(0^-)=2g_4 \ell \, \psi'(0^+)\,,&& \psi'(0^+)=\psi'(0^-)
 \end{align} 
 and is realized by the parameters
\begin{align}\label{eq:g4}
\eta=\frac{\pi}{2}+\arctan(g_4)\,,&&\m=\biggl(\frac{g_4}{\sqrt{1+g_4^2}}, \frac{1}{\sqrt{1+g_4^2}},0,0\biggr)\,.
\end{align}
 Notice that in the literature this interaction has sometimes been called, rather confusingly, $\ddelta'$-interaction, see e.g.~\cite{Lan15}.

\section{Generalized Dirac comb}
\label{sec:genComb}

\begin{figure}[t]
\centering
\includegraphics{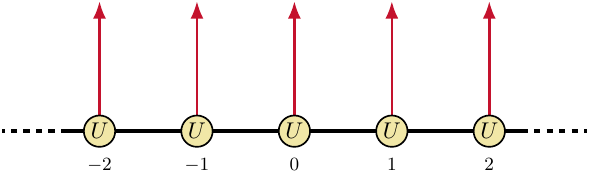}

\caption{Schematic representation of a generalized Dirac comb. At each lattice point $n\in\Z$ there is a generalized point interaction imposing the $\UU(2)$ boundary conditions $(I-U)\Psi_n'=\iu (I+U)\Psi_n$.}
\label{fig:comb}
\end{figure}

We now consider a \emph{generalized Dirac comb}, that is, a periodic array  of generalized point interactions on the real line, with lattice spacing $L$. We describe each point interaction in terms of the boundary conditions associated with a  $2\times 2 $  unitary matrix $U\in\UU(2)$, as explained in the previous section. By choosing $\ell=L$ as the unit of length,  and $\hbar^2/(2mL^2)$ as the unit of energy, the Hamiltonian  just reads
\begin{equation}\label{eq:HU}
H_U=-\frac{ \dd^2}{\dd x^2}\,,
\end{equation}
whereas its domain is
\begin{equation}
\dom(H_U)=\bigl\{\psi\in\Sob^2(\R\!\setminus\!\Z): (I-U)\Psi_n'=\iu (I+U)\Psi_n\,, \forall n \in \Z \bigr\}
\end{equation}
with $\Z$ being the lattice of point interactions and
\begin{align}  
\Psi_n=\mat{  \psi(n^-)\\  \psi(n^+)} \,, \qquad \Psi_n'= \mat{ \psi'(n^-) \\ -\psi'(n^+)} \,.
\end{align}
Therefore, the equation 
 $(I-U)\Psi_n'=\iu (I+U)\Psi_n$
imposes boundary conditions for $\psi(x)$ and its derivative, that are the same at each lattice point  $n\in\Z$, see Fig.~\ref{fig:comb}. Also in this case, we can realize $H_U$ by using suitable point-interactions. By exploiting the Kurasov mapping, in the rescaled variables of Eq.~\eqref{eq:HU} we  obtain the expression
\begin{align}\label{eq:Vg}
    \tilde{H}_{\g }=-\frac{ \dd^2}{\dd x^2}+\sum_{n\in\Z}V_{\g }(x-n)\,,
\end{align}
with $\g=\g(U)$ as in~\eqref{eq:gU}, and where the singular potential $V_{\g }$ in the rescaled units reads
\begin{equation}\label{eq:vg}
    \frac{V_{\vb g}(x)}{2}=g_1 \ddelta(x)+g_2 \ddelta'(x)+\iu  g_3  \biggl(\frac{\dd}{\dd x}\ddelta(x)+\ddelta(x)\frac{\dd}{\dd x}\biggr)+g_4 \frac{\dd}{\dd x}\ddelta(x)\frac{\dd}{\dd x}\,.
\end{equation}

The operator $H_U$ is a self-adjoint extension of the second derivative operator initially defined on $C_0^\infty(\R\!\setminus\!\Z)$, which preserves both the topology of the real line and the periodicity of the lattice. We mention that, in principle, one could consider far more general self-adjoint extensions (that is, more general boundary conditions) by waiving one of these two requests. By moving to the setting of (infinite) quantum graphs, where each point of the lattice is now understood as a vertex of a certain metric graph, one can change the topology of the system: a ``loop'' can be described by considering non-local boundary conditions which effectively identify two or more points of the lattice $\Z$.
On the other hand, by attaching to each point of the lattice a different local boundary condition $U_n$, one can keep the topology of the line but obtains a non-periodic self-adjoint extension. 

A very interesting system also involving an infinite series of $\ddelta$-like potentials on the real line has been proposed as a solution of the conjecture raised by P\'{o}lya and Hilbert, that the Riemann zeros are the eigenvalues of a quantum mechanical Hamiltonian. The system  describes the quantum dynamics of a massless Dirac fermion moving in a region of Rindler spacetime under the action of delta function potentials localized on the square free integers. The corresponding Hamiltonian admits a self-adjoint extension that is tuned to the phase of the zeta function, on the critical line, in order to obtain the Riemann zeros as bound states. The model suggests a possible
proof of the Riemann hypothesis in the limit where the singular potentials vanish~\cite{GS1,GS2}.

In the following, after exploiting in section~\ref{sec:bloch} the Bloch transform to decompose the periodic Hamiltonian $H_U$ in a fibered operator on the Brillouin zone, in
section~\ref{sec:spectral} we take advantage of this decomposition to discuss some spectral properties of $H_U$, deriving in particular its spectral function. In
section~\ref{sec:valence} we then use the latter to determine which  $\ddelta$-like point-interactions admit a valence energy band associated with localized states.

\subsection{Bloch decomposition}\label{sec:bloch}
Let us recall that the Bloch-Floquet transform is defined for any $\psi\in\mathcal{S}(\R)$ by~\cite{ReSi78, BeSh91, Pa07, PaPi13, MoPaPi18, DeNLe11}
\begin{align}
(\UB\psi)(y,k)= \frac{1}{\sqrt{2\pi}}\sum_{n\in \Z} \e^{-\iu k n} \psi(y+n) \,,
\end{align}
where $y\in [-\frac{1}{2},\frac{1}{2})$ and $k\in \mathcal{B}=[-\pi,\pi)$, and extended uniquely to a unitary operator
\begin{align}
\UB\colon L^2(\R)&\to \mathcal{H}_\text{B}=\int_{\mathcal{B}}^\oplus\Leb^2_k(-\tfrac{1}{2},\tfrac{1}{2}) \,\dd k	\,.
\end{align}
In the above equations $\mathcal{B}$  denotes the Brillouin zone, $k$ is the dimensionless quasimomentum  whereas $\mathcal{H}_\text{B}$ is the direct integral Hilbert space with fiber
\begin{align}\label{eq:L2k}
\Leb^2_k(-\tfrac{1}{2},\tfrac{1}{2})\cong \Leb^2(-\tfrac{1}{2},\tfrac{1}{2})
\end{align}
and inner product  
\begin{align}
\langle\phi_1|\phi_2\rangle_{\mathcal{H}_\text{B}}=\int_{\mathcal{B}} \langle\phi_1(\cdot,k)|\phi_2(\cdot,k)\rangle_{\Leb^2_k(-\frac{1}{2},\frac{1}{2}) }\,\dd k\,.
\label{eq:innprod}
\end{align}
The Bloch transformed wave function $\phi(y,k)=(\mathcal{U}_\text{B}\psi)(y,k)$, when extended as a wave function of $\Leb^2_\text{loc}(\R)\otimes \Leb^2_\text{loc}(\R)$, is pseudo-periodic in the real space and periodic in the reciprocal space, that is
\begin{align}\label{eq:periodicity}
\phi(y+n,k)=\e^{\iu kn } \phi(y,k)\,,&& \phi(y,k+2\pi n)=\phi(y,k)\,,
\end{align}
for any $n\in \Z$ and for all $y,k\in\R$. This means that quasimomenta are defined \emph{modulo} $2\pi$, and that the Brillouin zone can be identified as the set of the equivalence classes, that is as  the one-dimension torus $\R/2\pi\Z$. Henceforth we will not use this identification explicitly, allowing the quasimomenta to take any real value but considering, when needed, the following representative of $k\in\R$ in the fundamental cell $[-\pi,\pi)$,
\begin{equation}\label{eq:modulo}
[k]\coloneq( k+\pi \bmod 2\pi) -\pi =
k-2\pi\Bigl \lfloor \frac{k-\pi}{2\pi}\Bigr\rfloor \,,
\end{equation}
where $\lfloor x \rfloor$ denotes the floor of $x$, i.e.\ the largest integer less or equal to $x$~\cite{Knuth72}.\footnote{An alternative approach to deal with the periodicity of the Brillouin zone is to consider the mapping $\chi\colon k\mapsto \e^{\iu k}$ from the quasimomentum to its \emph{Floquet multiplier}, see e.g.~\cite{DeNLe11}.}
We also recall that the inverse Bloch transform is given by
\begin{equation}
(\UB^\dagger\phi)(x)=\frac{1}{\sqrt{2\pi}}\int_{\mathcal{B}} \e^{\iu kn_x}\phi(y_x,k) \,\dd{k}\, ,
\end{equation}
 where $(y_x,n_x)\in [-\tfrac{1}{2},\tfrac{1}{2})\times \Z $ is the unique pair such that $x=y_x+n_x$.

By exploiting the Bloch transform, the Hamiltonian $H_U$ is unitarily equivalent to the fibered operator
\begin{align}
\hat{H}_U\coloneq\UB H_U \UB^\dagger =\int_{\mathcal{B}}^\oplus h_{U}(k)\,\dk\,,
\end{align}
where the fiber Hamiltonian,
\begin{equation}\label{eq:hUk}
    h_U(k)=-\frac{ \dd^2}{\dd y^2}\,,
\end{equation}
acts on the  Hilbert space $L^2_k(-\tfrac{1}{2},\tfrac{1}{2})$ in~\eqref{eq:L2k},
with domain contained in
\begin{align}\label{eq:H2k}
H^2_k=\bigl\{\psi\in H^2\bigl([-\tfrac{1}{2},\tfrac{1}{2}]\!\setminus\!\{0\}\bigr): \psi^{(j)}\bigl(\tfrac{1}{2}\bigr)=\e^{\iu k}\psi^{(j)}\bigl(-\tfrac{1}{2}\bigr),\, j=0,1 \bigr\},
\end{align}
namely the space of wave functions satisfying the  Bloch pseudo-periodic boundary conditions at the boundary of the unit cell. The domain of $h_U(k)$ is further restricted by the boundary conditions $U$ coming from the periodic array of point interactions, which now are applied just at the point $y=0$:
\begin{equation}\label{eq:domhU}
\dom(h_U(k))=\{ \psi\in H^2_k: (I-U)\Psi_0'=\iu (I+U)\Psi_0\}.
\end{equation}

\begin{figure}[t]
\centering
\includegraphics{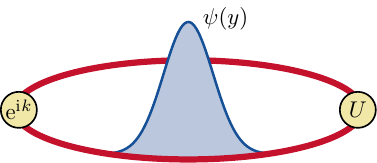}

\caption{The physical system described by the fiber Hamiltonian $h_U(k)$ is a ring with two generalized point interactions at antipodal points. In this figure, the left and right point interactions implement respectively the Bloch condition 
$\psi^{(j)}(1/2)=\e^{\iu k}\psi^{(j)}(-1/2)$ with $j=0,1$ and  the $\UU(2)$ condition
$(I-U)\Psi_0'=\iu (I+U)\Psi_0$.} 
\label{fig:ring}
\end{figure}

The fiber operator $h_U(k)$ has an interesting physical interpretation on its own. It represents indeed the Hamiltonian of a particle in a ring, of radius $1/(2\pi)$, interacting with two point-like defects, or junctions, located at antipodal points, see Fig.~\ref{fig:ring}. In this picture, one junction realizes the $\UU(2)$ boundary condition $U$ and the other the $\UU(1)$  pseudo-periodic Bloch condition. As is well known~\cite{AIM05}, two point defects on a ring could accommodate a larger $\UU(2)\times \UU(2)$ family of local boundary conditions, or even a $\UU(4)$ family if we allow for topology-changing boundary conditions mapping the ring into a figure-eight shaped curve. The system that we want to study represents thus a particular instance of this more general model, associated with a $\UU(1)\times \UU(2)$ subfamily. We mention that a different $\UU(1)\times \UU(2)$ subfamily has been considered in~\cite{parities}, where a generic $\UU(2)$ interaction at one point of the ring was combined with the $\UU(1)$ family of Robin conditions at the other point, effectively changing the topology of the ring to that of a box. On the other hand, the Hamiltonian $h_U(0)$ describing a particle in a ring with a single $\UU(2)$ point interaction has been extensively investigated in~\cite{isoboundary}.

\subsection{Spectral function}\label{sec:spectral} 
From general results, see e.g.~\cite{ReSi78}, we know that the spectrum of  $H_U$ has a band structure
\begin{equation}  
\spec(H_U)=\spec(\hat{H}_U) =  \overline{\bigcup_{n\in\N } \{\epsilon_{U,n}(k):k \in [-\pi,\pi) \}} \, ,
\end{equation}
where $\epsilon_{U,n}(k)$ denotes the $n$-th eigenvalue of $h_U(k)$ and the overline denotes the closure of the set.  In other words, solving the spectral problem for $H_U$ is equivalent to solve the eigenvalue problem for the fiber Hamiltonian $h_U(k)$ in~\eqref{eq:hUk} for each $k\in\mathcal{B}$.
When $\epsilon_U(k)\neq 0$, in particular, a solution of the differential equation 
\begin{equation}
    -\frac{\partial^2}{\partial y^2}\phi(y,k)=\epsilon_U(k)\phi(y,k) \, ,
\end{equation}
that takes into account the pseudo-periodic condition~\eqref{eq:H2k} at the boundary of the cell is given by
\begin{equation}\label{eq:psik}
\phi_\epsilon(y,k)=\begin{cases}
c_1 \e^{\iu qy}+c_2 \e^{-\iu qy}\,,& y\in[-1/2,0)\\
c_1 \e^{\iu[q(y-1)+k]} +c_2\e^{-\iu[q(y-1)-k]} \,,& y\in(0,1/2]
\end{cases}
\end{equation}
where 
\begin{equation}\label{eq:q}
    q=q(\epsilon)=\e^{\iu \arg(\epsilon)/2}\sqrt{|\epsilon|}
\end{equation}
and $\epsilon=\epsilon_U(k)$; when clear from the context, we will omit the dependence of $\epsilon_U(k)$ from $U$ and $k$. Notice that, although the dimensionless energy $\epsilon$ is always real, the dimensionless wavenumber $q$  can be either real or purely imaginary. 

To further impose the $\UU(2)$ boundary conditions~\eqref{eq:domhU} at $y=0$, we proceed as in~\cite{isoboundary}:  by inserting the above solution into the quantities 
\begin{equation}
     \Psi_{k,\pm}(\epsilon) =\mat{+\phi_\epsilon'(0^-,k)\pm\iu \phi_\epsilon(0^-,k)\\ -\phi_\epsilon'(0^+,k)\pm\iu \phi_\epsilon(0^+,k)}
\end{equation}
we can obtain the matrices $A_{k,\pm}(\epsilon)$ defined by
\begin{equation}
	\Psi_{k,\pm}(\epsilon) =A_{k,\pm}(\epsilon) \mat{c_1\\c_2},
\end{equation}
so that imposing the boundary condition in~\eqref{eq:domhU}, that can be now rewritten as $\Psi_{k,-}(\epsilon) =U\Psi_{k,+}(\epsilon) $, is equivalent to requiring that
\begin{equation} 
F_{U,k}(\epsilon)=\det \bigl(B_k(\epsilon)-U\bigr)=0\,,
\end{equation}
where $B_k(\epsilon)= A_{k,-}(\epsilon) A^{-1}_{k,+}(\epsilon)$. For the latter quantity we find the expression
\begin{align}\label{eq:Bk}
B_k(\epsilon)=a(\epsilon)I+b(\epsilon)\sigma_k \, ,
\end{align}
where 
\begin{align}\label{eq:sigmak}
\sigma_k=\cos(k)\sigma_x+\sin(k)\sigma_y=\mat{0 & \e^{-\iu k} \\ \e^{\iu k} & 0}
\end{align}
and
\begin{align}
    a(\epsilon)=\frac{(q^2-1)\sin(q)}{(q^2+1)\sin(q)-2\iu q \cos(q)}\,,&& 
    b(\epsilon)=\frac{2\iu q}{(q^2+1)\sin(q)-2\iu q \cos(q)},
\end{align}
with $q=q(\epsilon)$ as in~\eqref{eq:q}.
We conclude that the energy bands, and hence the whole spectrum of $H_U$, can be found by determining the real zeroes of the \emph{spectral function} $F_{U,k}(\epsilon)$, that is
\begin{equation}\label{eq:sigmaHU2}
    \spec(H_U)=\overline{\bigcup_{k\in[-\pi,\pi)}\ker_{\R} F_{U,k}(\epsilon)}\,.
\end{equation}

We should remark here that,  when $\epsilon_U(k)=0$, the structure of the eigenfunctions $\psi_0(x,k)$ is quite different from that of Eq.~\eqref{eq:psik}, as in this case they would be given by piecewise linear functions. In principle, thus, the condition $F_{U,k}(0)=0$ would be inconclusive to decide whether $\epsilon_{U}(k)=0$ is actually an eigenvalue or not. Notwithstanding, a careful computation on the lines of that given in~\cite{isoboundary}, that we omit for brevity, reveals that $B_k(\epsilon)$ is continuous at $\epsilon=0$ and that its limit for $\epsilon\to 0$ coincides with the  quantity $B_{k}(0)=A_{k,-}(0)A_{k,+}^{-1}(0)$ constructed by starting from the actual eigenfunctions $\psi_{0}(x,k)$ associated with $\epsilon=0$, hence allowing us to confirm the validity of Eq.~\eqref{eq:sigmaHU2} for any $\epsilon\in\R$.

We now derive the explicit expression of the spectral function. Since for any two $2\times 2$ matrices $M$ and $N$ we have that
\begin{align}
\det(M-N)=\det(M)+\det(N)+\tr(MN)-\tr(M)\tr(N)\,,
\end{align}
by using the decomposition in~\eqref{eq:Bk} and the fact that $\tr(\sigma_k)=0$ we find that
\begin{align}\label{eq:FUk0}
F_{U,k}(\epsilon)=a(\epsilon)^2-b(\epsilon)^2+\det(U)-a(\epsilon)\tr(U)+b(\epsilon)\tr(U\sigma_k)\,.
\end{align}
Then, by dropping a non-vanishing multiplicative factor and using the constraint $m_0^2+m_1^2+m_2^2+m_3^2=1$, we arrive at the explicit expression
\begin{align}\label{eq:FUk1}
F_{U,k}(\epsilon)&=m_1 \cos(k)+m_2\sin(k)-G_{\eta,m_0}(\epsilon), 
\end{align} 
where
\begin{gather}
G_{\eta,m_0}(\epsilon)=\frac{1}{2} \sinc(q)\bigl[q^2\bigl(\cos(\eta)-m_0\bigr)+\cos(\eta)+m_0\bigr]+\cos(q)\sin(\eta),
\label{eq:FUk2}
\end{gather}
with $\sinc(q)=\sin(q)/q$. Interestingly, the spectral function fully depends on $\eta$ and $m_0$ but it does not depend on $m_3$, and it has a “residual” dependence on $m_1$ and $m_2$ only through the  $k$-dependent linear combination  $m_1 \cos(k)+m_2\sin(k)$. This signals the presence of isospectrality, which will be extensively investigated in section~\ref{sec:isospectrality}. For $k=0$, in particular, we recover the case already studied in~\cite{isoboundary}. 

Before proceeding, we point out two simple particular cases.
For the pseudo-periodic conditions~\eqref{eq:Upp}, that are obtained for $\eta=\pi/2$, $m_0=m_3=0$, $m_1=\cos(\alpha)$ and $m_2=\sin(\alpha)$, the spectral function reduces to
\begin{align}
F_{U,k}(\epsilon)=\cos(k-\alpha)-\cos(q)\,.
\end{align}
In this case the energy bands span the whole positive axis and there are no gaps, that is $\operatorname{spec}(H_U)=[0,+\infty)$. This is a consequence of the fact that the generalized Dirac comb with pseudo-periodic conditions with phase $\alpha$ is unitarily equivalent to the free Hamiltonian on the real line (that, in turn, can be though of as a generalized Dirac comb with periodic conditions)  via the unitary ``gauge'' transformation
\begin{equation}
    \psi(x)\mapsto \e^{-\iu \alpha \sum_{n\in\Z}\Uptheta(x-n)}\psi(x) \, ,
\end{equation}
where $\Uptheta(x)$ is the Heaviside step function.

For the asymmetric Robin conditions~\eqref{eq:URobin},  that is for $m_1=m_2=0$, the spectral function~\eqref{eq:FUk1} becomes instead independent of $k$ and all the energy bands are flat. This should come with no surprise since the Robin conditions are confining and they decouple the line in an infinite number of disjoint intervals, not allowing for the transmission between adjacent intervals. 

Although a comprehensive discussion of the transport properties for the other generalized Dirac combs is beyond the scopes of this article, we conclude this section by determining which singular interactions admit the presence of valence energy bands.

\subsection{Valence bands}\label{sec:valence} 
For our purposes the valence bands are energy bands in the lowest part of the energy spectrum, characterized by negative energies and by states whose wave functions are localized around the singular point of the lattice. 
The  negative energies are linked to the existence of pure imaginary zeroes of the spectral function, i.e.  solutions of the equation $F_{U,k}(-\mathfrak{q}^2)=0$  where we set $\epsilon=(\iu \mathfrak{q})^2$, replacing $q\mapsto \iu\mathfrak{q}$, with $\mathfrak{q}\in \R$.  In  particular, this replacement only affects the part of $F_{U,k}(\epsilon)$ in~\eqref{eq:FUk1} which  does not depend on $k$, namely the term in~\eqref{eq:FUk2}:
\begin{equation}
G_{\eta,m_0}(-\mathfrak{q}^2)=\frac{\sinh(\mathfrak{q})}{2\mathfrak{q}}\left(\mathfrak{q}^2(m_0-\cos(\eta))+m_0+\cos(\eta)\right)+\cosh(\mathfrak{q})\sin(\eta).
\end{equation}

\begin{figure}[tb]
\centering
\includegraphics{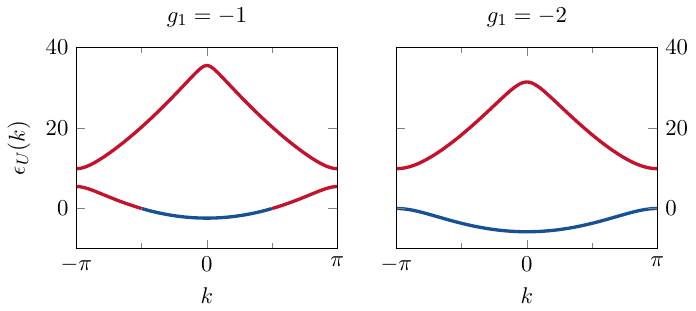}
 
	\caption{The first two energy bands for the $\ddelta$-potentials $\g=(g_1,0,0,0)$ with $g_1=-1$ (left) and $g_1=-2$ (right). The positive energies are shown in red and the negative ones in blue. }\label{fig:delta_bands}

\end{figure}

We now analyze the existence of this kind of solutions for the  four different families of point-like interactions described in section~\ref{sec:pointint}. 
 Let us start with the case of $\ddelta$-potentials, associated with the couplings $\g=(g_1,0,0,0)$. 
 Using the parametrization given by Eq.~\eqref{eq:g1} the spectral function reduces to
\begin{equation}
F_{U,k}(-\mathfrak{q}^2)=g_1\sinh(\mathfrak{q})+\mathfrak{q}\bigl(\cosh(\mathfrak{q})-\cos(k)\bigr).
\end{equation}
There are three different regimes depending on the values of $g_1$. 
When $g_1>0$  there are no localized states for any value of $k$. When $-2< g_1<0$  there are solutions but only for some values of $k$, and this means that the lowest energy band contains both negative and positive energies (see the left panel in  figure~\ref{fig:delta_bands}).  When $g_1<-2$ there is a solution for all $k$ and the lower energy band is fully made of negative energy states (see the right panel in figure~\ref{fig:delta_bands}). We remark that for this singular interaction there is always an energy gap between the different energy bands, for any $g_1\in \R$.

For what concerns the $\ddelta'$-potentials, i.e. $\g=(0,g_2,0,0)$, by using Eq.~\eqref{eq:g2} we arrive at the spectral function
\begin{equation}
	F_{U,k}(-\mathfrak{q}^2)=\cosh(\mathfrak{q})-\frac{1-g_2^2}{1+g_2^2}\cos(k).
\end{equation}
In this case there are no solutions (except at $\mathfrak{q}=0$) for any value of $k$, and therefore there are no localized states. We mention that if there is a combination of the couplings $g_1$ and $g_2$,  the existence of negative energy states depends on the relative strength of the two couplings. When $g_2>g_1$, for example, there are no localized states (see~\cite{Spectrum-Cas} for an extensive analysis of this hybrid situation).

Moving to the singular gauge field interaction
$v_3(x)$ in~\eqref{eq:v3v4def}, corresponding to   $\g=(0,0,g_3,0)$, the spectral function takes the form
\begin{equation}
	F_{U,k}(-\mathfrak{q}^2)=\cosh(\mathfrak{q})-\cos(\alpha-k),
\end{equation}
with $g_3=\tan(\alpha/2)$. As in the previous case, there are no negative energy states since the only allowed solution is $\mathfrak{q}=0$. This is consistent with the fact that a $\g=(0,0,g_3,0)$ interaction coincides with a generalized Dirac comb with the pseudo-periodic boundary conditions in Eq.~\eqref{eq:Upp}, and we have already observed that the latter is unitarily equivalent to the free Hamiltonian.

\begin{figure}[tb]
\centering
\includegraphics{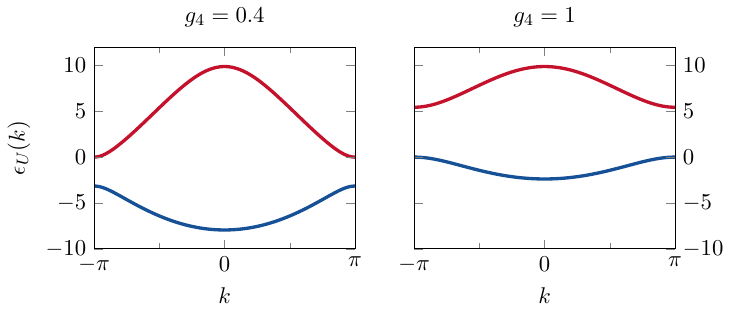}
 
\caption{The first two energy bands for the singular metric $\frac{\dd{}}{\dd x}\ddelta(x)\frac{ \dd{} }{\dd x}$ interactions $\vb{g}=(0,0,0,g_4)$ with $g_4=0.4$ (left) and $g_4=1$ (right). The positive energies are shown in red and the negative ones in blue.}
\label{fig:delta4_bands}

\end{figure}

Finally, for the singular metric interaction
$v_4(x)$ in~\eqref{eq:v3v4def}, corresponding to  $\g=(0,0,0,g_4)$,   by
using Eq.~\eqref{eq:g4} we obtain the spectral function
\begin{equation}
F_{U,k}(-\mathfrak{q}^2)=g_4 \mathfrak{q} \sinh(\mathfrak{q})-\cosh(\mathfrak{q})-\cos (k).
\end{equation}
In this case there are also three different regimens depending on the value of $g_4$. If $g_4\le 0$  there are no negative energy states. If $0<g_4< 1/2$  there are solutions 
for every value of $k$ and the valence band is full of negative energy states which do not reach the zero energy at the end of the band, creating a gap between the valence band and the first conducting band, see the left panel in figure~\ref{fig:delta4_bands}. If $g_4> 1/2$ the negative energy band reaches the zero energy but the first conducting band starts from a positive energy, so there is still a gap between the bands, see the right panel in figure~\ref{fig:delta4_bands}. The only case when the gap closes is for the critical value $g_4=1/2$, and for any $g_4\neq 1/2$ there is not a mixed band with negative and positive energy like what happens for the $\ddelta$-potential.

\section{Isospectrality}\label{sec:isospectrality}
The general expression in Eq.~\eqref{eq:FUk0} shows that the spectral function $F_{U,k}(\epsilon)$, and \textit{a fortiori} the spectrum $\operatorname{spec}(H_U)$, depend on the boundary unitary matrix $U$ only through  $\det(U)$, $\tr(U)$ and $\tr(U\sigma_k)$, with $\sigma_k$ given in~\eqref{eq:sigmak}. As a matter of fact the latter three independent quantities cannot  fully “reconstruct” the space of boundary conditions, that is the four-dimensional group $\UU(2)$. Therefore, we expect to find a set of \emph{isospectral} maps from $\UU(2)$ to itself preserving the spectrum of the system, in a sense that we are now going to carefully explicate. Let us first notice that translations in $k$ of the spectral function, that we may denote as \emph{horizontal} transformations of the latter, do not affect the global spectrum of $H_U$, 
since $F_{U,k}(\epsilon)$ is $2\pi$-periodic in $k$ and
\begin{equation}
    \overline{\bigcup_{k\in[-\pi,\pi)}\ker_{\R} F_{U,k}(\epsilon)}=\overline{\bigcup_{k\in[-\pi,\pi)}\ker_{\R} F_{U,f(k) }(\epsilon)}
\end{equation}
for any bijection $f\colon [-\pi,\pi)\to [-\pi,\pi)$. %
A particular example, which will be considered later, is a rigid translation of the form $f(k)=[k+\delta]$ with $\delta\in\R$, see Eq.~\eqref{eq:modulo}. In general, the displacement 
\begin{equation}
\delta(k)=f(k)-k	
\end{equation}
 will have a non-trivial dependence on the quasimomentum $k$.

We conclude that to any map $U\mapsto \phi_k(U)\in \UU(2)$ that preserves the spectral function up to a displacement in $k$, that is realizing the equality
\begin{equation}\label{eq:FUtransf}
F_{U,f(k)}(\epsilon)=F_{\phi_k(U),k}(\epsilon)
\end{equation}
for all $k\in\mathcal{B}$ and for a certain bijection $f$ on $\mathcal{B}$, there corresponds a pair of Hamiltonians
$H_U$ and $\int_\mathcal{B}^\oplus h_{\phi_k(U)}(k)\,\dd k$ which are isospectral, i.e.\ related by
\begin{equation} \label{eq:isospec}
\spec(H_U)=\spec\Bigl( \int_\mathcal{B}^\oplus h_{\phi_k(U)}(k)\,\dd k \Bigr)\,.
\end{equation}

For later convenience let us introduce the following distinction: we say that a transformation of the spectral function as in~\eqref{eq:FUtransf} is \emph{vertical} if $f$ is the identity map, i.e.\ $\delta(k)\equiv0$, and is \emph{oblique} if otherwise $\delta(k)\neq 0$, that is if it combines a vertical and a horizontal transformation.

In the rest of this section, after discussing in Sec.~\ref{sec:Fsymm} all the relevant ``boundary symmetries'' of the spectral function, namely the transformations of the form~\eqref{eq:FUtransf}, in Sec.~\ref{sec:lifting} we determine the corresponding isospectrality relations between pairs of Hamiltonians.

\subsection{Boundary symmetries of the spectral function}\label{sec:Fsymm}

For practical reasons, and following~\cite{isoboundary,  parities, isodirac}, in this article we restrict our considerations to  maps on $\UU(2)$ which are  smooth bijections preserving the $\UU(2)$ group structure, that is we will only consider maps of the form $U\mapsto \phi_k(U)$ where $\phi_{k}$ is an element of $\Aut^\pm(\UU(2))$, the group of automorphisms and anti-automorphisms of $\UU(2)$. See Appendix~\ref{app:auto} for more details. 

We mention that by considering only smooth bijections $\phi_{k}\colon \UU(2)\to \UU(2)$, without requiring the maps $\phi_{k}$ also preserve the group structure,  we would  arrive at $\mathrm{Diff}(\UU(2))$, the infinite-dimensional group of diffeomorphisms of $\UU(2)$: a careful analysis of the latter is beyond the scopes of the present article, and will be object of future investigations.

On the other hand, $\Aut^\pm(\UU(2))$ turns out to be a finite-dimensional (Lie) group, that is explicitly  characterized by the semi\-direct product
\begin{align}
\Aut^\pm(\UU(2))\cong \Inn(\UU(2))\rtimes \{\id,\kappa,\iota,\tau \} \, ,
\end{align}
where $\Inn(\UU(2))\cong \SU(2)/\Z_2=\SU(2)/\{I,-I\}$ is the subgroup of inner automorphisms of $\UU(2)$ whereas, besides the identity map $\id(U)=U$, $\kappa(U)=\overline{U}$ is the complex conjugate matrix of $U$, $\iota(U)=U^{-1}=U^\dagger$ is the inverse matrix of $U$ (equal to its adjoint, as $U$ is unitary), and $\tau(U)=(\kappa\circ\iota)(U)=U^\intercal$ is the transpose matrix of $U$. In particular, $\kappa$ is an outer automorphism, whereas both $\iota$ and $\tau$ are anti-automorphisms. The elements of $\Inn(\UU(2))$ are the $\UU(2)$ conjugations
\begin{align}
\phi_{\nu(\delta,\vb{n})}\colon U\mapsto \nu(\delta,\vb{n}) U \nu(\delta,\vb{n})^\dagger,
\end{align}
which are in one-to-one correspondence with the elements
\begin{align}
\nu(\delta,\vb{n})=\e^{-\iu \frac{\delta}{2} (n_x\sigma_x+n_y\sigma_y+n_z\sigma_z)}\in  \SU(2)/\Z_2
\end{align}
where $\delta\in[0, \pi]$ and $\vb{n}=(n_x,n_y,n_z)$ with $\|\vb{n}\|^2=1$.

By observing that for a generic $U\in \UU(2)$ and for any $\nu\in \SU(2)/\Z_2$ 
\begin{subequations}
\begin{gather}
\det(\nu U\nu^\dagger)=\det (U)\,,\\
\det( U^\intercal)=\det(U)\,, \\
\det( \overline{U})=\det( U^\dagger)=\overline{\det(U)} \neq \det(U)\,,
\end{gather}
\end{subequations}
with analogous relations holding for the trace, we conclude that only inner automorphisms and the anti-automorphisms associated with $\tau$ preserve the quantities $\det(U)$ and $\tr(U)$, and are thus relevant for our analysis. What remains, in order to determine all the isospectral maps $\phi_{k}$ realizing the identity \eqref{eq:FUtransf}, is to investigate which ones further preserve the quantity $\tr(U\sigma_k)$  with $\sigma_k$ given in~\eqref{eq:sigmak}, in the sense that
\begin{equation}\label{eq:trsigmak}
    \tr(\phi_{k}(U)\sigma_k)=\tr(U\sigma_{f(k)})
\end{equation}
for a certain bijection $f\colon \mathcal{B}\to \mathcal{B}$. In the following we discuss separately the inner automorphisms and the anti-automorphisms.

\subsubsection{Inner automorphisms}
We first consider vertical transformations, i.e.\ transformations which do not alter the value of $k$, satisfying $\tr(\phi_{k}(U)\sigma_k)=\tr(U\sigma_k)$. We deduce that the only admissible conjugations are given by
\begin{align}\label{eq:vertU}
U\mapsto U^{\delta}_\text{ver}(k)=\e^{-\iu \frac{\delta(k)}{2}\sigma_k}U\e^{\iu \frac{\delta(k)}{2}\sigma_k}\,,
\end{align}
with $\delta(k)\in \R$ and $\sigma_k$ as in~\eqref{eq:sigmak},
that is for any function $\delta\colon \mathcal{B}\to \R$ we find the following $k$-dependent vertical transformation of the spectral function:
\begin{align}\label{eq:FUvert}
    F_{U^{\delta}_\text{ver}(k),k}(\epsilon)= F_{U,k}(\epsilon)\,.
\end{align}
For later convenience let us notice that the only $k$-independent fixed points of Eq.~\eqref{eq:vertU} are given by symmetric Robin conditions:
\begin{align}
\{\e^{\iu\alpha}I:\alpha\in[0,2\pi)\}\,.
\end{align}

Moving to oblique transformations, that can shift the value of $k$, we find that the only admissible conjugations are given by
\begin{align}\label{eq:oblU}
U\mapsto U^{\delta}_{\text{obl}}(k)=\e^{-\iu\frac{\delta(k)}{2}\sigma_z} U\e^{\iu\frac{\delta(k)}{2}\sigma_z}\,, 
\end{align}
for any non-trivial function  $\delta\colon \mathcal{B}\to\R$. Setting  $f(k)=[k+\delta(k)]$, if $f$  is a bijection on $\mathcal{\mathcal{B}}$ then
\begin{align}\label{eq:FUobl}
    F_{U_{\text{obl}}^\delta(k),k}(\epsilon) = F_{U,f(k)}(\epsilon),
\end{align}
is an oblique transformation of the spectral function. In this case, the fixed points form a two-parameter family corresponding with the asymmetric Robin conditions~\eqref{eq:URobin}, namely
\begin{align}\label{eq:UO}
\{\e^{\iu(\alpha I+\beta \sigma_z)}: \alpha\in[0,\pi)\,,\beta\in[0,2\pi)\}\,.
\end{align}
As it turns out, thus, the fixed points of the oblique transformations include the $k$-independent fixed points of the vertical transformations.

\subsubsection{Anti-automorphisms}
For what concerns the anti-automorphisms associated with $\tau$, namely the maps
\begin{align}
    U\mapsto \nu U^\intercal \nu^\dagger\,, \qquad \nu\in  \SU(2)/\Z_2\,,
\end{align}
by observing that
\begin{align}
\tr(\nu U^\intercal \nu^\dagger \sigma_k)=\tr(U^\intercal \nu^\dagger \sigma_k\nu )=\tr\bigl(U  \nu^\intercal \sigma_{k}^\intercal(\nu^\intercal)^\dagger  \bigr)\,.
\end{align}
and that $\sigma_k^\intercal=\sigma_{-k}=\sigma_x\sigma_{k}\sigma_x$, we  conclude that in this case the allowed vertical transformations are obtained by selecting
\begin{align}
\nu \in \{ \iu (\e^{\iu\frac{\delta}{2} \sigma_k}\sigma_x)^\intercal=\iu \e^{\iu\frac{\delta}{2} \sigma_k}\sigma_x:\delta\in[0,2\pi) \}\,,
\end{align}
which gives the family of anti-automorphisms
\begin{align}\label{eq:antivert}
U\mapsto \e^{\iu\frac{\delta(k)}{2} \sigma_k}\sigma_x U^\intercal \sigma_x \e^{-\iu\frac{\delta(k)}{2} \sigma_k}\,, 
\end{align}
whereas the oblique transformations are obtained by selecting
\begin{align}
\nu\in\{\iu(\e^{\iu\frac{\delta}{2} \sigma_z}\sigma_x)^\intercal=
\iu \e^{-\iu\frac{\delta}{2} \sigma_z}\sigma_x
: \delta \in[0,2\pi)\}\,,
\end{align} 
which gives the family
\begin{align}\label{eq:antiobl}
U\mapsto \e^{-\iu\frac{\delta(k)}{2} \sigma_z}\sigma_x U^\intercal \sigma_x \e^{\iu\frac{\delta(k)}{2} \sigma_z}\,. 
\end{align}
Notice that the discrete anti-automorphism
\begin{equation}
U\mapsto \sigma_x U^\intercal \sigma_x
\label{eq:antiauto}	
\end{equation}
is obtained from $U$ by letting $m_3\mapsto -m_3$, see Eq.~\eqref{eq:U}. Therefore, the anti-automorphisms in Eqs.~\eqref{eq:antivert} and~\eqref{eq:antiobl} combine the discrete transformation~\eqref{eq:antiauto} with the continuous transformations provided by the inner automorphisms in Eqs.~\eqref{eq:vertU} and~\eqref{eq:oblU}, respectively.

\subsection{Bulk-edge correspondence}\label{sec:lifting}
In the previous section we found that the spectral function possesses a set of \emph{boundary} symmetries, each  with a  corresponding isospectrality relation via Eq.~\eqref{eq:isospec}, that are associated with two continuous transformations given by the inner automorphisms  in Eqs.~\eqref{eq:vertU} and~\eqref{eq:oblU}, possibly combined with a discrete transformation associated with the anti-automorphism $U\mapsto \sigma_x U^\intercal \sigma_x$, see Eqs.~\eqref{eq:antivert} and~\eqref{eq:antiobl}. Now, we want to show that each boundary symmetry can be also lifted to a \emph{bulk} symmetry, so that the Hamiltonians are not only isospectral but even \emph{unitarily equivalent} (or antiunitarily equivalent, in the case of an anti-automorphism). Our analysis completes the  results presented in~\cite{ExGr99,KuLa02, CheShi04}, offering a unified framework to discuss all the isospectrality relations.

\subsubsection{Vertical unitary  transformations}
Given an arbitrary displacement function $\delta\colon \mathcal{B}\to \R$, the vertical transformation in Eq.~\eqref{eq:vertU} can be induced by the unitary fiber operator 
\begin{align}
 u_{\text{ver}}^\delta(k) =\e^{\iu\frac{\delta(k)}{2}P_k} 
 \,,&& P_k=\cos(k) P_x+\sin(k) P_y
 \label{eq:vertudelta}
\end{align}
acting on the fiber Hilbert space $\Leb^2_k(-\frac{1}{2},\frac{1}{2})$, where
\begin{align}\label{eq:PxPy}
    (P_x\psi)(y)=\psi(-y)\,,&&(P_y\psi)(y)=\iu\sign(y)\psi(-y)\,,
\end{align} 
with $y\in[-\frac{1}{2},\frac{1}{2})$,
are two \emph{generalized parities}~\cite{parities}. Although individually $P_x$ and $P_y$ are not fiber-preserving, since  for any $\psi\in H^2_k$ we have that $P_x\psi\in H^2_{-k}$ and $P_y\psi\in H^2_{[\pi-k]}$, their combination appearing in $P_k$ is, i.e.\ we have that $P_k\psi\in H^2_k$ for any $\psi\in H^2_k$.

\begin{figure}[t]
\centering
\includegraphics{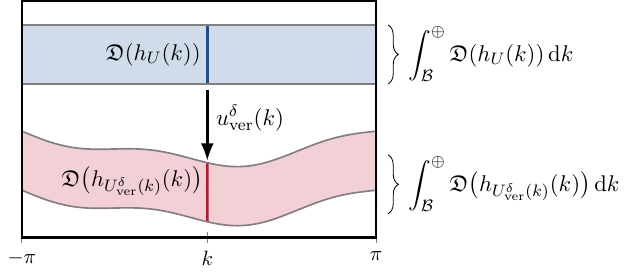}
 
\caption{Pictorial representation of a vertical unitary transformation. Notice how,  ignoring the Bloch pseudo-periodicity, the $k$-independent original domain (in blue) is mapped to a $k$-depending domain (in red), and the corresponding bulk Hamiltonian can no longer be interpreted as a (local) generalized Dirac comb.}
\label{fig:ver}
\end{figure}

As the reader can easily verify,  $u_{\text{ver}}^\delta(k)$ does not modify the bulk expression~\eqref{eq:hUk} of $h_U(k)$, 
but changes its domain~\eqref{eq:domhU} according to
\begin{align}
u_{\text{ver}}^\delta(k) \dom(h_U(k))=\dom\bigl(h_{U_{\text{ver}}^\delta(k)}(k)\bigr)\,,
\end{align}
as sketched in Fig.~\ref{fig:ver}. 
The global fibered unitary operator  on  $\mathcal{H}_{\mathcal{B}}$ reads
\begin{align}
\hat{\mathcal{U}}_{\text{ver}}^\delta=
\int_{\mathcal{B}}^\oplus   u_{\text{ver}}^\delta(k) \, \dd k \,,
\end{align}
with $u_{\text{ver}}^\delta(k)$ as in~\eqref{eq:vertudelta}.
Therefore, since
\begin{align}
\hat{\mathcal{U}}_{\text{ver}}^\delta \hat{H}_U
\hat{\mathcal{U}}_{\text{ver}}^{\delta\dagger} 
&=\hat{\mathcal{U}}_{\text{ver}}^\delta
\left(\int_{\mathcal{B}}^\oplus h_U(k) \dd k \right)
\hat{\mathcal{U}}_{\text{ver}}^{\delta\dagger} \nonumber
= \int_{\mathcal{B}}^\oplus   u_{\text{ver}}^\delta(k) h_U(k) u_{\text{ver}}^\delta(k)^\dagger\, \dd k
\nonumber\\
&= \int_{\mathcal{B}}^\oplus h_{U^{\delta}_\text{ver}(k)}(k)\,\dd k  \,,
\end{align}
the operator ${\mathcal{U}}_{\text{ver}}^\delta=\UB^\dagger\hat{\mathcal{U}}_{\text{ver}}^\delta\UB$ unitarily maps $H_U$ to the transformed Hamiltonian
\begin{align}\label{eq:HV}
{\mathcal{U}}_{\text{ver}}^\delta 
H_U {\mathcal{U}}_{\text{ver}}^{\delta\dagger} =\UB^\dagger
\left(\int_{\mathcal{B}}^\oplus  h_{U_\text{ver}^\delta(k)}(k) {\dd k}\right)
\UB\, ,
\end{align} 
with 
$U^{\delta}_\text{ver}(k)=\e^{-\iu \frac{\delta(k)}{2}\sigma_k}U\e^{\iu \frac{\delta(k)}{2}\sigma_k}$ as in~\eqref{eq:vertU}.

\begin{figure}[t]
\centering
\includegraphics{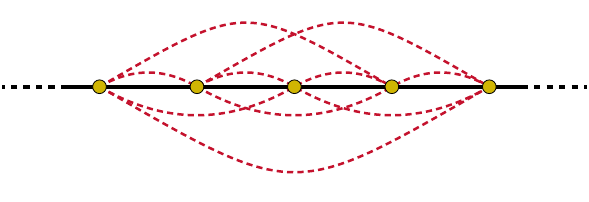}

\caption{Schematic representation of the non-local model described by Eq.~\eqref{eq:nonlocal} when $V(x)$ is supported only at the lattice points. The dashed red lines show the non-local interactions between all the lattice points.}
\label{fig:nonlocal}
\end{figure}

Since the domain of $ h_{U_\text{ver}^\delta (k)}(k) $ has an additional dependence on $k$ induced by $U_\text{ver}^\delta (k)$, beside the Bloch pseudo-periodicity, the operator in~\eqref{eq:HV}   in general contains interactions which are non-local in real space, and is thus no longer a generalized Dirac comb. Notice that this is true even if the displacement function is independent of $k$, i.e.\ if $\delta(k)=\delta$.  To better understand the physical interpretation of the right-hand side of Eq.~\eqref{eq:HV}, and how the $k$-depending boundary conditions of the fiber Hamiltonian $h_{U_\text{ver}^\delta(k)}(k)$ are associated with non-local effects, we recall that, by having in mind the Kurasov mapping, $h_{U_\text{ver}^\delta(k)}(k)$ can be realized by a free fiber Hamiltonian with a generalized point interaction with $k$-depending couplings, say $\g=\g(k)$. Let us thus consider a generic fiber Hamiltonian of the form
\begin{equation}
h(k)=-\frac{ \dd^2}{\dd y^2}+g(k) V(y),
\end{equation}
for $y\in [-\frac{1}{2}, \frac{1}{2})$,
where $V(x)$ is a periodic 
potential of period $1$. Then the action of 
\begin{equation}
H=\UB^\dagger \left(
\int_{\mathcal{B}}^\oplus   h(k) \dd k
\right)\UB
\end{equation}
on a suitable $\psi(x)\in \mathcal{S}(\R)$, setting $x=y+n$ with $y\in [-\tfrac{1}{2},\tfrac{1}{2})$ and $n\in\Z$, is
\begin{align}
(H\psi)(x) 
&=\frac{1}{\sqrt{2\pi}}\int_{\mathcal{B}}\e^{\iu k n} 
\bigl(h(k) \UB\psi\bigr)(y,k) \,\dd k \nonumber\\
&=\frac{1}{2\pi}\int_{\mathcal{B}}\e^{\iu k n} \sum_{m\in \Z} \e^{-\iu  k m} 
\left(-\frac{\dd^2}{\dd y^2}+g(k) V(y)\right)
\psi (y+m)\,\dd k\nonumber\\
&= \sum_{m\in \Z} \frac{1}{2\pi}\int_{ \mathcal{B}}\e^{\iu (n-m)  k} \biggl(-\frac{ \dd^2\psi}{\dd y^2}(y+m)+g(k) V(y)\psi(y+m)\biggr) \,\dd k \nonumber\\
&=-\frac{ \dd^2\psi}{\dd x^2}(x)+ V(x-n)\sum_{m\in \Z} \hat{g}_{n-m} \psi(x-(n-m)) 
\end{align}
where 
\begin{equation}
\hat{g}_{n}=\frac{1}{2\pi}\int_{ \mathcal{B}}\e^{\iu   k n} g(k) \,\dd{ k}\,
\end{equation}
are the Fourier coefficients of the coupling.
By using the periodicity of the potential we obtain the final expression
\begin{align}
(H\psi)(x) =-\frac{ \dd^2}{\dd x^2}\psi(x)+\sum_{n\in \Z} \hat{g}_{n} V(x)\psi(x-n) ,
\label{eq:nonlocal}
\end{align}
which clearly shows the non-local behavior of $H$. In the particular case in which $V(x)$ is a singular potential supported only at the lattice points $x\in\Z$, as it happens for the Kurasov potential, the action of $H$ is non-local only at the lattice points, remaining local in the bulk $\R\!\setminus\!\Z$ (see Fig.~\ref{fig:nonlocal}).

\subsubsection{Oblique unitary transformations}
The oblique transformation in Eq.~\eqref{eq:FUobl} cannot be induced by a bulk unitary operator which is fibered, because its action in general maps a fiber onto a different one.

\begin{figure}[t]
\centering
\includegraphics{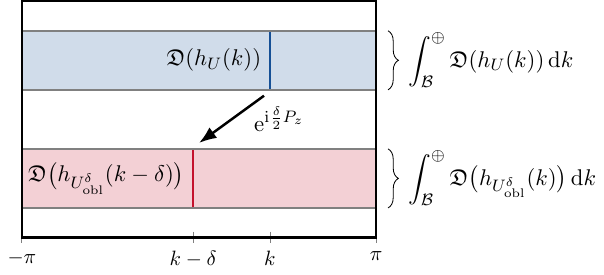}
 
\caption{Pictorial representation of the oblique unitary transformation. In this case the transformed domain (in red) has no additional dependence on $k$, beside the Bloch pseudo-periodicity, and the corresponding bulk Hamiltonian is the generalized Dirac comb with boundary condition $U_\text{obl}^\delta$.}
\label{fig:obl}
\end{figure}

On the unit cell Hilbert space $L^2(-\frac{1}{2},\frac{1}{2})$ consider the generalized parity
\begin{align}\label{eq:Pz}
 (P_z\psi)(y)=\iu(P_yP_x\psi)(y)=-\sign(y)\psi(y)
\end{align}
and the action of the unitary it generates,
\begin{align}
 \psi_\delta(y)=\bigl(\e^{\iu\frac{\delta}{2}P_z} \psi\bigr)(y) =  \e^{-\iu\frac{\delta}{2}\sign(y)} \psi(y)\, .
\end{align}
If $\psi$ satisfies the Bloch pseudo-periodic boundary conditions~\eqref{eq:H2k} with quasimomentum $k$, that is $\psi\in H^2_k$, then $\psi_\delta$ will have boundary conditions with quasimomentum $k'=k-\delta$, that is $\psi_\delta\in H^2_{ [k-\delta]}$.
On the other hand, the boundary condition $U$ in~\eqref{eq:domhU} satisfied by $\psi$ at $y=0$ will go into $U^{\delta}_{\text{obl}}$ in~\eqref{eq:oblU}. 
Therefore, we have that the unitary generated by $P_z$ acts on the domain~\eqref{eq:domhU} of the kinetic energy Hamiltonian as
\begin{equation}
 \e^{\iu\frac{\delta}{2}P_z} \dom\bigl(h_U(k)\bigr)=\dom\bigl(h_{U_\text{obl}^\delta}(k-\delta)\bigr)\,,
\end{equation}
while leaving its action unchanged, see Fig.~\ref{fig:obl}.

We can therefore construct the following unitary on the total space $\mathcal{H}_B$. Consider a bijective and (almost everywhere) absolutely continuous function $f\colon\mathcal{B}\to \mathcal{B}$ describing the $k$-dependent horizontal shift $\delta(k)=f(k)-k$. Then, the operator $\hat{\mathcal{U}}_\text{obl}^\delta$  which acts on $\phi\in\mathcal{H}_B$ as
\begin{equation}
	(\hat{\mathcal{U}}_\text{obl}^\delta \phi)(y,k) = \sqrt{|f'(k)|} \, \e^{-\iu\frac{f(k)-k}{2}\sign(y)} \phi\bigl(y,f(k)\bigr),
\end{equation}
for $y\in [-\frac{1}{2},\frac{1}{2})$ and $k\in \mathcal{B}$, is unitary. Indeed, by Eq.~\eqref{eq:innprod} we have that
\begin{align}
	\|\hat{\mathcal{U}}_\text{obl}^\delta \phi\|^2_{\mathcal{H_B}} &= \int_{\mathcal{B}} \|\hat{\mathcal{U}}_\text{obl}^\delta \phi(\cdot,k)\|^2_{\Leb^2(-\frac{1}{2},\frac{1}{2}) }\,\dd k
	\nonumber\\
	&= \int_{\mathcal{B}} \|\phi(\cdot,f(k))\|^2_{\Leb^2(-\frac{1}{2},\frac{1}{2}) }
	|f'(k)|
	\,\dd k 
	= \|\phi\|^2_{\mathcal{H_B}}\,, 
\end{align}
for all $\phi\in\mathcal{H}_B$. Moreover,
\begin{align}
\bigl(\hat{\mathcal{U}}_\text{obl}^\delta \hat{H}_U
\hat{\mathcal{U}}_\text{obl}^{\delta\dagger} \phi\bigr)(y,k) &= \sqrt{|f'(k)|} \, \e^{-\iu\frac{\delta(k)}{2}\sign(y)} \bigl(\hat{H}_U
\hat{\mathcal{U}}_\text{obl}^{\delta\dagger} \phi\bigr)\bigl(y,f(k)\bigr)
\nonumber\\
&=\sqrt{|f'(k)|} \, \e^{-\iu\frac{\delta(k)}{2}\sign(y)} h_U\bigl(f(k)\bigr)\bigl(
\hat{\mathcal{U}}_\text{obl}^{\delta\dagger} \phi\bigr)\bigl(y,f(k)\bigr)
\nonumber\\
&= \e^{-\iu\frac{\delta(k)}{2}\sign(y)} h_U\bigl(f(k)\bigr) \e^{\iu\frac{\delta(k)}{2}\sign(y)}\phi(y,k)
\nonumber\\
&=h_{U_\text{obl}^\delta(k) }(k) \phi(y,k)\,, 
\end{align}
that is
\begin{align}
\hat{\mathcal{U}}_\text{obl}^\delta \hat{H}_U
\hat{\mathcal{U}}_\text{obl}^{\delta\dagger} 
=\int_{\mathcal{B}}^\oplus h_{U_\text{obl}^\delta(k) }(k) \,\dd k\,.
\end{align}

Therefore, the operator ${\mathcal{U}}_{\text{obl}}^\delta=\UB^\dagger\hat{\mathcal{U}}_{\text{obl}}^\delta\UB$ unitarily maps $H_U$ to the transformed Hamiltonian
\begin{align}\label{eq:HOb}
{\mathcal{U}}_{\text{obl}}^\delta 
H_U {\mathcal{U}}_{\text{obl}}^{\delta\dagger}
=\UB^\dagger
\left(\int_{\mathcal{B}}^\oplus  h_{U_\text{obl}^\delta(k)}(k) {\dd k}\right)
\UB\,.
\end{align} 
This has the same form of the Hamiltonian ${\mathcal{U}}_{\text{ver}}^\delta 
H_U {\mathcal{U}}_{\text{ver}}^{\delta\dagger}$ obtained in~\eqref{eq:HV} by a vertical transformation. Again, the dependence of the boundary conditions $U_\text{obl}^\delta (k)$ on the quasimomentum $k$ gives rise to non-local interactions, so that the transformed Hamiltonian~\eqref{eq:HOb} is no longer a generalized Dirac comb, as discussed in the previous section.

However, at variance with vertical transformations, now there are non-trivial situations where the boundary conditions $U_\text{obl}^\delta (k)$ is the same for all the fibers. This happens if the horizontal displacement $\delta (k)$ is (almost everywhere) independent of the quasimomentum $k$, i.e.\ if $[\delta (k)]=\delta$, which corresponds to the bijection $f\colon \mathcal{B}\to \mathcal{B}$ with
\begin{equation}
  f(k) =[ k+ \delta] \,.
\end{equation}
In such situation, the boundary unitary $U_\text{obl}^\delta$ is independent of $k$, and one obtains a unitary equivalence between the two Dirac combs with boundary conditions $U$ and $U_\text{obl}^\delta$:
\begin{align}\label{eq:HUoblring}
{\mathcal{U}}_{\text{obl}}^\delta H_U {\mathcal{U}}_{\text{obl}}^{\delta\dagger}=H_{U^\delta_\text{obl}}\, ,
\end{align}
where
\begin{align}\label{eq:oblUconst}
U^{\delta}_{\text{obl}}=\e^{-\iu\frac{\delta}{2}\sigma_z} U\e^{\iu\frac{\delta}{2}\sigma_z}\,. 
\end{align}

We remark that this relation is non-trivial for all the boundary conditions that are different from the family of fixed points in~\eqref{eq:UO}, that is for all non-Robin conditions. In other words, by starting from any generalized Dirac comb associated with a non-Robin boundary condition $U$, by means of Eq.~\eqref{eq:HUoblring} we can construct a $\UU(1)$ family $\{ H_{U^\delta_\text{obl}} \}_{\delta\in [0,2\pi)}$ of generalized Dirac combs such that, for any $\delta\in(0,2\pi)$ we have that $U\neq U^\delta_\text{obl}$ and that $H_U$ is unitarily equivalent, and hence isospectral, to $H_{U^\delta_\text{obl}}$.

\subsubsection{Antiunitary transformations}
 To conclude our analysis, we observe that the discrete boundary symmetry
 \begin{equation}
     F_{\sigma_x U^\intercal \sigma_x,k}(\epsilon)=F_{U,k}(\epsilon)
 \end{equation}
associated with the anti-automorphism $U\mapsto \sigma_x U^\intercal \sigma_x$ can be lifted to a bulk symmetry by considering the  antiunitary  operator $P_x T$, where
\begin{equation}
  (T\phi)(y,k)= \overline{\phi(y,-k)}
\end{equation}
 is the antiunitary  operator of \emph{time-reversal} and the parity operator $P_x$ is defined in~\eqref{eq:PxPy}, so that $(P_xT \phi) (y,k)=\overline{\phi(-y,k)}$ is fiber-preserving and
 \begin{align}
P_xT \dom(h_U(k))=\dom(h_{\sigma_xU^\intercal \sigma_x}(k))\,.
\end{align}
By mapping the global fibered operator on $\mathcal{H}_B$  generated by  $P_x T$ to $L^2(\R)$,  one gets
\begin{equation}
	{\mathcal{V}}_{P_xT} =  \UB^\dagger \left(\int_{\mathcal{B}}^\oplus   P_x T \, \dd k  \right)\UB \,
\end{equation}
that establishes an antiunitary equivalence
\begin{equation}
{\mathcal{V}}_{P_xT} H_U {\mathcal{V}}_{P_xT}^\dagger
=H_{\sigma_x U^\intercal \sigma_x}
\label{eq:VPT}
\end{equation}
between two generally different Dirac combs. Notice that, as for unitary transformations, also antiunitary transformations preserve the spectrum, that is we have the isospectrality relation
\begin{equation}
    \spec(H_{\sigma_x U^\intercal \sigma_x})= \spec(H_U)\,,
\end{equation}
for any $U\in\UU(2)$. Finally, the isospectrality relations associated with the anti-automorphisms in Eqs.~\eqref{eq:vertU} and~\eqref{eq:oblU} can be easily obtained by composing the action of the antiunitary operator ${\mathcal{V}}_{P_xT}$ in~\eqref{eq:VPT} with that of the vertical and oblique unitary operators $\mathcal{U}_\text{vert}^\delta$ in~\eqref{eq:HV} and $\mathcal{U}_\text{obl}^\delta$ in~\eqref{eq:HOb}, respectively.

\section{Discussion and outlook}\label{sec:discussion}
In this work we have developed a comprehensive framework to study the isospectrality problem for a generalized Dirac comb, formalizing the analysis of \cite{ExGr99, FulTsu00, parities, KuLa02, CheShi04, isoboundary, isodirac} and extending their results. By characterizing the energy spectrum of a Dirac comb $H_U$ in terms of a spectral function, we managed to fully analyze a group of transformations preserving the latter. Taking advantage of the Bloch decomposition, we  determined a large family of isospectral maps, thereby identifying which fiber Hamiltonians $h_U(k)$ share the same spectrum up to (eventually) a shift in the quasimomentum. More in detail, we found three families of isospectral maps: two continuous inner automorphisms, given by the vertical and oblique maps in Eqs.~\eqref{eq:vertU} and~\eqref{eq:oblU}, and a set of anti-automorphisms combining the latter with the discrete anti-automorphism given in Eq.~\eqref{eq:antiauto}. 
For each isospectral map, that connects two boundary conditions $U$ and $U'$, we were also able to construct a fiber unitary (or antiunitary) operator connecting the corresponding fiber Hamiltonians, and to lift the latter to a  unitary (or antiunitary) operator acting on the bulk Hilbert space $L^2(\R)$. Moreover, we identified specific regimes in the parameter space where valence bands composed of localized states with negative energy appear. These findings can have applications in the engineering of band structures and wave localization in artificial quantum materials or metamaterials, where externally-controlled defects can be used to tune spectral features.

The oblique automorphisms with rigid horizontal translation and the discrete anti-auto\-mor\-phism, in particular, turned out to be independent of $k$  and to give respectively rise to unitary and antiunitary operators connecting two generally distinct Dirac combs $H_U$ and $H_{U'}$. As  derived in Eq.~\eqref{eq:UO}, the only fixed points of the oblique transformations are the asymmetric Robin conditions. Among these, only the subset of \emph{symmetric} Robin conditions are also invariant with respect to the discrete anti-automorphism. The latter is indeed essentially related to a parity transformation, and any Dirac comb with non-symmetric Robin condition is trivially unitarily equivalent to its mirror image. This means that all generalized Dirac combs with point interactions which are \emph{not} associated with Robin conditions, i.e.\ all and only the non-confining point interactions, are not uniquely determined by their spectrum and thus \emph{cannot be heard}, using the words of Kac~\cite{Kac}. More precisely, any Dirac comb $H_U$ with non-confining interactions is unitarily equivalent, via Eq.~\eqref{eq:HUoblring}, to the non-trivial $\UU(1)$ family $\{H_{U^\delta_{\text{obl}}} \}_{\delta\in[0,2\pi)}$ of other distinct Dirac combs. Our analysis, however, does not implies that Dirac combs with symmetric Robin conditions are uniquely determined from their spectrum, as in principle there could exist other isospectral maps connecting them, beside the ones we considered (for instance, a diffeomorphism of $\UU(2)$ not preserving its group structure). Nevertheless, from a direct inspection of their spectral function we can easily check that Dirac combs with symmetric Robin conditions are spectrally unique, and constitute thus the only generalized Dirac combs that can be heard.

Vertical isomorphisms and nonrigid oblique isomorphisms, on the other hand, map a $k$-independent condition $U$ to a condition $U^\delta_{\text{ver}}(k)$ or $U^\delta_{\text{obl}}(k)$ depending on the quasimomentum. As we explained, the corresponding bulk unitary operators map then generalized Dirac combs to operators that are non-local at the lattice points, see Eq.~\eqref{eq:nonlocal}.
Even tough the physical interpretation of the latter is still unclear, this result is very interesting when read in the opposite direction: we found that a particular family of non-local periodic operators is unitarily equivalent to generalized Dirac combs, which are clearly easier to describe and  characterize.

Looking ahead, there are several interesting directions for future research that are related to our analysis. One natural extension is the study of quasi-periodic or disordered arrangements of point interactions~\cite{Jona, DrKiSB12, Frohlich}, where the spectral theory becomes more intricate and localization phenomena can arise. Another possible extension is the investigation of periodic system involving a particle with spin, eventually considering the relativistic counterpart of the Kronig--Penney model involving the Dirac equation~\cite{AvrGro76, CaMaPo13}.  Additionally, it would be interesting to explore the topological aspects of the isospectral structure uncovered here, potentially linking our results to the well-known classification of topological phases in one-dimensional systems~\cite{CaFu21}. Finally, although our work focuses on the one-dimensional case, many of the ideas and techniques developed here may be adapted to higher-dimensional systems, at least when there exists an explicit solution of the differential problem at hand.

\section*{Acknowledgments}
We thank Beppe Marmo for useful discussions. 
GA acknowledges support from MUR-Italian Ministry of University and Research and Next Generation EU within PRIN 2022AKRC5P ``Interacting Quantum Systems: Topological Phenomena and Effective Theories'' and from the Italian National Group of Mathematical Physics (GNFM-INdAM).
PF acknowledges support from INFN through the project ``QUANTUM'', from the Italian National Group of Mathematical Physics (GNFM-INdAM), from  PNRR MUR project CN00000013 - ``Italian National Centre on HPC, Big Data and Quantum Computing'', and from the Italian funding within the ``Budget MUR - Dipartimenti di Eccellenza 2023--2027''  - Quantum Sensing and Modelling for One-Health (QuaSiModO).
The work of M.A. and F.F. is partially supported  by Spanish MINECO/FEDER Grants No. PGC2022-126078NB-C21 funded by MCIN/AEI/ 10.13039/501100011033, ERDF A way of making Europe Grant; the Quantum Spain project of the QUANTUM ENIA of  {\sl Ministerio de Asuntos Econ\'omicos y Transformaci\'on Digital},  {\sl Diputaci\'on General de Arag\'on Fondo Social Europeo (DGA-FSE)} Grant No. 2020-E21-17R of Aragon Government, and {\sl Plan de Recuperaci\'on, Transformaci\'on y Resiliencia}- supported European Union – NextGenerationEU Program on {\sl Astrof{\'{\i}}sica y F{\'{\i}}sica de Altas Energ{\'{\i}}as}, CEFCA-CAPA-ITAINNOVA. 

\appendix
\section{Kurasov mapping}\label{app:Kurasov}
In this appendix we report the details of the Kurasov mapping connecting the Hamiltonians $H_{0,U}$ and $\tilde{H}_{0,\vb{g}}$ discussed in section \ref{sec:genpointint}. Let us first consider the $\UU(2)$ family of self-adjoint Hamiltonians 
\begin{align}
    H_{0,U}=-\frac{\hbar^2}{2m}\frac{\dd^2 }{\dd x^2}\,,&& U\in\UU(2)\,
\end{align}
defined in the domain
\begin{equation}
    \dom(H_{0,U})=\{\psi\in\Sob^2(\R\!\setminus\!\{0\}): (I-U)\Psi_0'=\iu (I+U)\Psi_0\}
\end{equation}
where 
\begin{align}
\Psi_0=\mat{\psi(0^-)\\ \psi(0^+)}\,, && \Psi'_0=\ell \mat{\psi'(0^-)\\ - \psi'(0^+)}\,.
\end{align}
 When the matrix $I-U$ is invertible, the boundary condition at $x=0$ can be also expressed as
\begin{align}
    \Psi'_0=\mathcal{C}(U)\Psi_0\,,&&\mathcal{C}(U)=\iu\frac{I+U}{I-U}\,,
\end{align}
$\mathcal{C}(U)$ being the inverse Cayley transform of $U$. In particular, by using the parametrization in Eq.~\eqref{eq:U}, we have that~\cite{echoes}
\begin{equation}
    \mathcal{C}(U)=\frac{1}{m_0-\cos(\eta)}\mat{-\sin(\eta)+m_3 & m_1-\iu m_2 \\ m_1+ \iu m_2 & -\sin(\eta)-m_3}\,.
\end{equation}

The boundary conditions $(I-U)\Psi_0'=\iu (I+U)\Psi_0$ can be also realized by adding to the kinetic-energy operator suitable point interactions supported at the origin $x=0$. By following~\cite{Kur96}, in particular, we have that the Kurasov Hamiltonian in Eq.~\eqref{eq:H0g}, that is\footnote{Notice that, at variance with the notations of~\cite{Kur96}, we put $\g=\frac{1}{2}(X_1,X_2,X_3,-X_4)$.} 
\begin{align} 
\tilde{H}_{0,\g}=-\frac{\hbar^2}{2m}\frac{ \dd^2}{\dd x^2}+g_1 v_1(x)+g_2 v_2(x)+g_3 v_3(x)+g_4 v_4(x)
\end{align}
where $\g=(g_1,g_2,g_3,g_4)$, acts as the free kinetic-term with domain
\begin{equation}
    \dom(\tilde{H}_{0,\vb g})=\{\psi\in\Sob^2(\R\!\setminus\!\{0\}): \Phi_1=\mathcal{M}_{\vb g}\Phi_2\}\,,
\end{equation}
where
\begin{align}
    \Phi_1=\mat{ \psi(0^+)\\ \ell\,\psi'(0^+)} \,,&& \Phi_2=\mat{ \psi(0^-)\\ \ell\,\psi'(0^-)} 
\end{align}
and
\begin{equation}
    \mathcal{M}_{\vb{g}}=\mat{ 
    \dfrac{(1+ g_2)^2+g_1g_4+g_3^2}{(1-\iu g_3)^2-g_1g_4-g_2^2} & \dfrac{2 g_4}{(1-\iu g_3)^2-g_1g_4-g_2^2} \\[10pt]
    \dfrac{2g_1}{(1-\iu g_3)^2-g_1g_4-g_2^2} & \dfrac{(1- g_2)^2+g_1g_4+g_3^2}{(1-\iu g_3)^2-g_1g_4-g_2^2}
    }\,.
\end{equation}
In order to match the two parametrizations, we need to express the components of $M_{\vb g}$ in terms of the components of $U$. We can achieve this by noticing that
\begin{align}
    \Phi_1=P_2\Psi_0-P_4 \Psi_0'\,,&&\Phi_2=P_1\Psi_0+P_3 \Psi_0'\,,
\end{align}
with
\begin{align}
    P_1=\mat{1 &0 \\ 0 & 0}\,,&&P_2=\mat{0 & 1 \\ 0 & 0}\,,&&P_3=\mat{0 & 0 \\ 1 & 0}\,,&&P_4=\mat{0 & 0 \\ 0 & 1}\,,
\end{align}
so that, by using $\Psi_0'=\mathcal{C}(U)\Psi_0$, we have that
\begin{align}
    \Phi_1=\mathcal{M}_{\vb g} \Phi_2 \quad \Leftrightarrow\quad (P_2-P_4\mathcal{C}(U))\Psi_0=M_{\vb g}(P_1+P_3\mathcal{C}(U))\Psi_0
\end{align}
and by solving the system 
\begin{equation}
    P_2-P_4\mathcal{C}(U)=\mathcal{M}_{\vb g}(P_1+P_3\mathcal{C}(U))
\end{equation}
we finally find the matching conditions given in Eq.~\eqref{eq:gU},
\begin{align}
\mat{g_1\\ g_2\\g_3\\g_4}=
\frac{1}{m_1+\sin(\eta)}\mat{m_0+\cos(\eta)\\-m_3\\m_2\\m_0-\cos(\eta))}\,.
\end{align}

To conclude let us notice that, according to~\cite{Lan15}, the Kurasov boundary conditions $\Phi_1=\mathcal{M}_{\vb g}\Phi_2$ can be naturally expressed as the jump-average boundary conditions introduced in Eq.~\eqref{eq:jumpaverageBC}, that is
\begin{align}
\mat{ \ell [\psi'(0)]\\[2pt] [\psi(0)]}=2M_{\vb g} \mat{ \{\psi(0)\}\\[2pt] \ell \{\psi'(0)\}}\,,&&M_{\vb g}=\mat{g_1& -g_2+\iu g_3\\ g_2+\iu g_3 &g_4}\,.
\end{align}

\section{Automorphisms and anti-automorphisms}\label{app:auto} 
In this appendix, after recalling  in section~\ref{sec:autgen} the definition (and some properties) of $\Aut^\pm(G)$, the group of automorphisms and anti-automorphisms of a given group $G$,  in section~\ref{sec:autuni} we discuss the particular case of the unitary group $G=\UU(2)$.  
\subsection{Definitions and general properties}\label{sec:autgen}
Let $G$ be a group, and let $\phi\colon G\to G$ be a bijective map. Then $\phi$ is a automorphism if
\begin{align}
\phi(gh)=\phi(g)\phi(h)\,.
\end{align}
The automorphisms of $G$ form a group with respect to composition, denoted by $\Aut(G)$. An inner automorphism is an automorphism induced by the conjugation action of $G$ on itself, that is a map of the form $\phi_g\colon h\mapsto ghg^{-1}$ for a given $g\in G$. The latter form a subgroup  $\Inn(G)$ of $\Aut(G)$, and one can prove that
\begin{align}
\Inn(G)\cong G/Z(G)\,,
\end{align}
where $Z(G)$ is the center of $G$, i.e.\ that we have the short exact sequence
\begin{align}
1\to Z(G)\to G\to \Inn(G)\to 1\,.
\end{align}
This characterization implies that $\Inn(G)$ is trivial if and only if $G$ is Abelian. It can also be proved that $\Inn(G)$ is a normal subgroup of $\Aut(G)$, that is we have the exact short sequence
\begin{align}\label{eq:IAOseq}
1\to \Inn(G)\to \Aut(G)\to \Out(G)\to 1
\end{align}
where the elements of the quotient group
\begin{align}
\Out(G)= \Aut(G)/\Inn(G)
\end{align}
are denoted as outer automorphisms. Let us recall that, in general, $\Out(G)$ cannot be identified with a subgroup of $\Aut(G)$, and thus non-inner automorphisms $\phi\in \Aut(G)\!\setminus\!\Inn(G)$ should be distinguished from outer automorphisms, which are actually left cosets of $\Inn(G)$ in $\Aut(G)$. This identification is however possible when the exact sequence in Eq.~\eqref{eq:IAOseq} splits, that is when there exists a section $s\colon \Out(G)\to \Aut(G)$, so that $\Aut(G)$ is given by the semidirect product $\Aut(G)\cong \Inn(G)\rtimes \Out(G)$ where the action of $\chi\in \Out(G)$ on $\phi\in\Inn(G)$ is $s(\chi)\circ\phi\circ s(\chi^{-1})$.\footnote{In this context, by denoting with $\pi$ the homomorphism from $\Aut(G)$ to $\Out(G)$, a section is a homomorphism $s\colon \Out(G)\to \Aut(G)$ such that $\pi\circ s =\id_{\Out(G)}$. }

We now move to anti-automorphisms~\cite{Rob96,EaNo16, Lia23}, that are bijective maps  $\rho\colon G\to G$ such that
\begin{align}
\rho(gh)=\rho(h)\rho(g)\,.
\end{align}
The inversion map 
\begin{align}
\iota_G\colon g\mapsto g^{-1}
\end{align}
is a particular anti-automorphism for any group $G$. Let $\Aut^-(G)$ denote the set of anti-automorphisms and $\Aut^\pm(G)=\Aut(G)\cup \Aut^-(G)$ the set of both automorphisms and anti-automorphisms. Since the composition of two anti-automorphisms is an automorphism whereas the composition of an anti-automorphism and of an automorphism is an anti-automorphism, the set $\Aut^-(G)$ is a group with respect to composition only if $G$ is Abelian, as in this case $\Aut^-(G)=\Aut(G)=\Aut^\pm(G)$. The set $\Aut^\pm(G)$ is instead always a group, and if $G$ is not Abelian we have the short exact sequence
\begin{align}
1\to \Aut(G)\to \Aut^\pm(G)\to \Z_2\to 1
\end{align}
which splits by the section mapping the non-trivial element of $\Z_2$ to $\iota_G$,
\begin{align}
\Aut^\pm(G)=\Aut(G)\rtimes \{\id_G, \iota_G\}\,.
\end{align}

\subsection{The unitary group}\label{sec:autuni}
Let us first recall that 
\begin{align}
\Inn(\UU(1))\cong 1\,, && \Aut(\UU(1))\cong \Out(\UU(1))=\Z_2
\end{align}
where the non-trivial element of $\Z_2$ is identified with the complex conjugation $\kappa_{\UU(1)}$, which in turn coincides with the inversion
\begin{align}
\kappa_{\UU(1)}(\e^{\iu\theta})=\overline{\e^{\iu\theta}}=\e^{-\iu\theta}=\iota_{\UU(1)}(\e^{\iu\theta})
\end{align}
as $\UU(1)$ is Abelian. Moving to $\SU(2)$, complex conjugation is no longer an outer automorphism, as it can be realized by conjugation with $\sigma_y$,
\begin{align}
\kappa_{\SU(2)}(\sigma)=\overline{\sigma}=\sigma_y\sigma\sigma_y 
\end{align}
and we thus have that
\begin{align}
 \Inn(\SU(2))\cong   \SU(2)/\Z_2\,, &&\Out(\SU(2))\cong1 
\end{align}
where $\Z_2=\{I,-I\}$, and 
\begin{align}
\Aut(\SU(2))=\Inn(\SU(2))\cong   \SU(2)/\Z_2\,.
\end{align}
In this case complex conjugation is related with the anti-automorphism $\tau_{\SU(2)}=\kappa_{\SU(2)}\circ\iota_{\SU(2)}$ given by transposition:
\begin{align}
\tau_{\SU(2)} (\sigma)=\overline{\sigma^{-1}}=\sigma^\intercal \,.
\end{align}
Collecting these results we conclude that for $\UU(2)$ 
\begin{align}
 \Inn(\UU(2))\cong  \UU(2)\UU(1)\cong \mathrm{PU}(2)\cong \SU(2)/\Z_2\,, 
\end{align}
where $\mathrm{PU}(2)$ denotes the projective unitary group,
\begin{align}\label{eq:U2out}
\Out(\UU(2))\cong\{\id_{\UU(2)}, \kappa_{\UU(2)} \}\cong \Z_2
\end{align}
and the sequence in Eq.~\eqref{eq:IAOseq} splits as
\begin{align}
\Aut(\UU(2))\cong ( \SU(2)/\Z_2)\rtimes \Z_2\,,
\end{align}
 with the action of $\Z_2$ on $\SU(2)/\Z_2$ given by Eq.~\eqref{eq:U2out}.  Also in this case the transposition $\tau_{\UU(2)}=\kappa_{\UU(2)}\circ\iota_{\UU(2)}$ is an anti-automorphism.

Summing up, we have that
\begin{align}
\Aut^\pm(\UU(2))\cong  ( \SU(2)/\Z_2)\rtimes K_4 \, ,
\end{align}
where
\begin{align}
K_4\cong \{\id_{\UU(2)},\kappa_{\UU(2)},\iota_{\UU(2)},\tau_{\UU(2)}\}\cong \Z_2\times \Z_2
\end{align}
is the Klein four-group. Therefore, we have the following four kind of transformations belonging to $\Aut^\pm(\UU(2))$: by denoting with $\sigma$ a generic element of $\SU(2)/\Z_2$ we have the inner 
 automorphisms 
\begin{align}
U\mapsto \sigma U\sigma^\dagger\,,
\end{align} 
the outer automorphisms
\begin{align}
U\mapsto \sigma \overline{U}\sigma^\dagger\,,
\end{align}
the anti-automorphisms of the first kind
\begin{align}
U\mapsto \sigma U^\intercal\sigma^\dagger
\end{align}
and the anti-automorphisms of the second kind 
\begin{align}
 U\mapsto \sigma U^\dagger \sigma^\dagger\,.
\end{align}
The terminology adopted for the anti-automorphisms follows~\cite{Alb39, KMRT98}: an anti-automorphism of $G$ is of the first kind (resp. of the second kind) if its restriction to $Z(G)$ is the identity map (resp. is an automorphism of period two, i.e.\ a non-trivial involution).

\end{document}